# Operando Raman Probing of Mode-Selective Electron–Phonon Coupling in Two-Dimensional Halide Perovskites

Tufan Paul[1], Helena Boom[2], Lilian Skokan[1], Viren Tyagi[2], Mostafa Shagar[1], Aditi Sahoo[1], Silvia Colella[3], Geert Brocks[2,4], Andreas Ruediger[1], Shuxia Tao[2], and Emanuele Orgiu [1]

[1] *Centre Énergie Matériaux Télécommunications, Institut national de la recherche scientifique, 1650 Blv. Lionel-Boulet, Varennes QC J3X 1P7, Canada.*

[2] *Department of Applied Physics and Science Education, Eindhoven University of Technology, 5600 MB Eindhoven, The Netherlands*

[3] *CNR NANOTEC, c/o Department of Chemistry, University of Bari, via Orabona 4, Bari 70125, Italy.*

[4] *Computational Chemical Physics, Faculty of Science and Technology and MESA+ Institute for Nanotechnology, University of Twente, 7500 AE Enschede, The Netherlands*

**Abstract**

Electron–phonon coupling governs charge transport, carrier relaxation, and polaron formation in halide perovskites, yet its microscopic origin in low-dimensional systems remains poorly understood. Here, we combine operando bias-dependent Raman spectroscopy with density functional theory (DFT) calculations to directly probe carrier–lattice interactions in two-dimensional Ruddlesden–Popper perovskites $(PEA)_2PbI_4$ and its fluorinated analogue $(PEA\text{-}F)_2PbI_4$. Under applied electric field, both systems exhibit mode-selective Raman linewidth broadening predominantly centered around ~100 $cm^{-1}$, while other phonon modes remain largely unaffected, revealing a highly selective coupling between injected carriers and specific lattice vibrations. DFT calculations identify these modes as hybrid organic–inorganic vibrations involving coupled motion of the organic spacer and symmetric Pb–I equatorial stretching, rather than purely inorganic phonons. Fluorination fundamentally reconstructs the vibrational landscape by altering molecular packing, crystal symmetry, and organic–inorganic coupling, leading to modified phonon density of states, longer phonon lifetimes, and enhanced carrier-mediated lattice response. Electrical bias drives opposite phonon lifetime evolution in thin films and single crystals. While the phonon lifetime decreases by ~17–22% in thin films, it increases by ~20–26% in single crystals, revealing that structural order fundamentally governs carrier–phonon interactions and phonon relaxation pathways in two-dimensional halide perovskites.

## 1. Introduction

Metal halide perovskites have emerged as an extremely promising and versatile class of semiconductors owing to their outstanding optoelectronic properties, including easy compositional/band-gap tunability, high absorption coefficients ($10^4$ to $10^5$ $cm^{-1}$ for visible light), balanced electron–hole diffusion lengths (>175 μm), and remarkable defect tolerance [1–5]. Three dimensional (3D) perovskites, with general formula $APbX_3$, have demonstrated unprecedented performance in photovoltaics and light-emitting devices; however, they often suffer of limited environmental and thermal stability due to their relatively soft lattice and susceptibility to moisture, heat, and ion migration [6, 7]. In contrast, two-dimensional (2D) halide perovskites with general formula $A_2PbX_4$, form naturally layered structures separated by organic layers, giving rise to quantum and dielectric confinement effects that strongly influence their electronic and optical behavior [8–12]. This layered architecture and the hydrophobic nature of the bulky organic cations provides enhanced environmental stability, reduced ion migration, and stronger quantum confinement effects, resulting in higher exciton binding energies and pronounced excitonic behavior [13, 14]. These features make low-dimensional perovskites attractive for applications such as light-emitting devices, lasers, and photodetectors, where they have demonstrated promising performances (EQE of 16.4% for light emitting diodes, responsivity 314 A $W^{-1}$, detectivity $3.4 \times 10^{13}$ Jones, and external quantum efficiency, EQE 86.5% for photodetectors) [15-18]. Their charge transport is generally more anisotropic and complex with respect to their 3D counterparts [12], where the inorganic $PbX_6$ octahedral framework plays a central role [19]. Subtle variations in octahedral geometry such as bond lengths, bond angles, and tilting distortions can significantly modify band structure, carrier mobility, and recombination dynamics [20–22]. At the same time, these structural motifs govern the lattice vibrational spectrum, linking atomic-scale distortions to phonon modes that directly influence thermal transport, carrier scattering, and energy dissipation [23, 24]. As a result, understanding how lattice vibrations couple to electronic degrees of freedom is essential for optimizing the performance of perovskite-based devices.

Electron–phonon (e–ph) coupling in halide perovskites is known to be remarkably strong due to the soft and polar nature of the lattice [25, 26]. In particular, low-frequency Pb–X (X= Cl, Br, I) vibrational modes and polar longitudinal optical (LO) phonons have been identified as key contributors to carrier scattering and linewidth broadening [27, 28]. Despite extensive studies on lattice dynamics in halide perovskites, most experimental approaches rely on temperature-dependent Raman spectroscopy or ultrafast optical techniques, which probe equilibrium or transient carrier dynamics, respectively [29, 30]. In contrast, direct probing of phonon behavior under an applied electric field i.e., under operando device conditions remains largely unexplored. Such an approach is essential to establish a direct link between lattice vibrations and charge transport in halide perovskite. The lack of operando investigations is particularly pronounced for 2D halide perovskites, where strong dielectric confinement and structural anisotropy are expected to substantially modify carrier–lattice coupling. Here, we address this gap by employing bias dependent Raman spectroscopy, enabling mode-resolved tracking of lattice dynamics during charge injection. This approach provides experimentally accessible insight into electron–phonon interactions without the need for large-scale facilities or ultrafast instrumentation, thereby offering a practical pathway to probe carrier–lattice coupling in halide perovskite devices. We further focus on fluorinated halide perovskites, which have emerged as a compelling materials platform owing to the strong electronegativity of fluorine that enables fine modulation of electronic structure [31]. Incorporation of fluorinated organic cations enhances environmental and thermal stability while suppressing defect-assisted recombination through tailored non-covalent interactions, rendering these materials particularly promising for next-generation optoelectronic and spintronic devices [32, 33].

Here, we employing *insitu* bias-dependent Raman spectroscopy to probe lattice dynamics in 2D halide perovskites, focusing on $(PEA)_2PbI_4$ and its fluorinated analogue $(PEA\text{-}F)_2PbI_4$. The experimental results are supported by phonon-mode analysis using a combination of density functional theory and machine-learning force-field calculations. By tracking the evolution of Raman-active phonon modes under applied electric fields, we identify a mode-selective response of the Pb–I stretching vibration, providing direct spectroscopic signatures of electron–phonon coupling within the inorganic framework. Furthermore, we compared experiments on single crystal and thin films, as representative of the close-to-ideal and close-to-application systems, getting further insights into how structural order and dielectric environment govern carrier–lattice interactions. These results demonstrate that bias-dependent Raman spectroscopy provides a practical route to probe and tune electron–phonon coupling in layered perovskites under device-relevant conditions.

## 2. Results

### Synthesis of two-dimensional (2D) perovskite

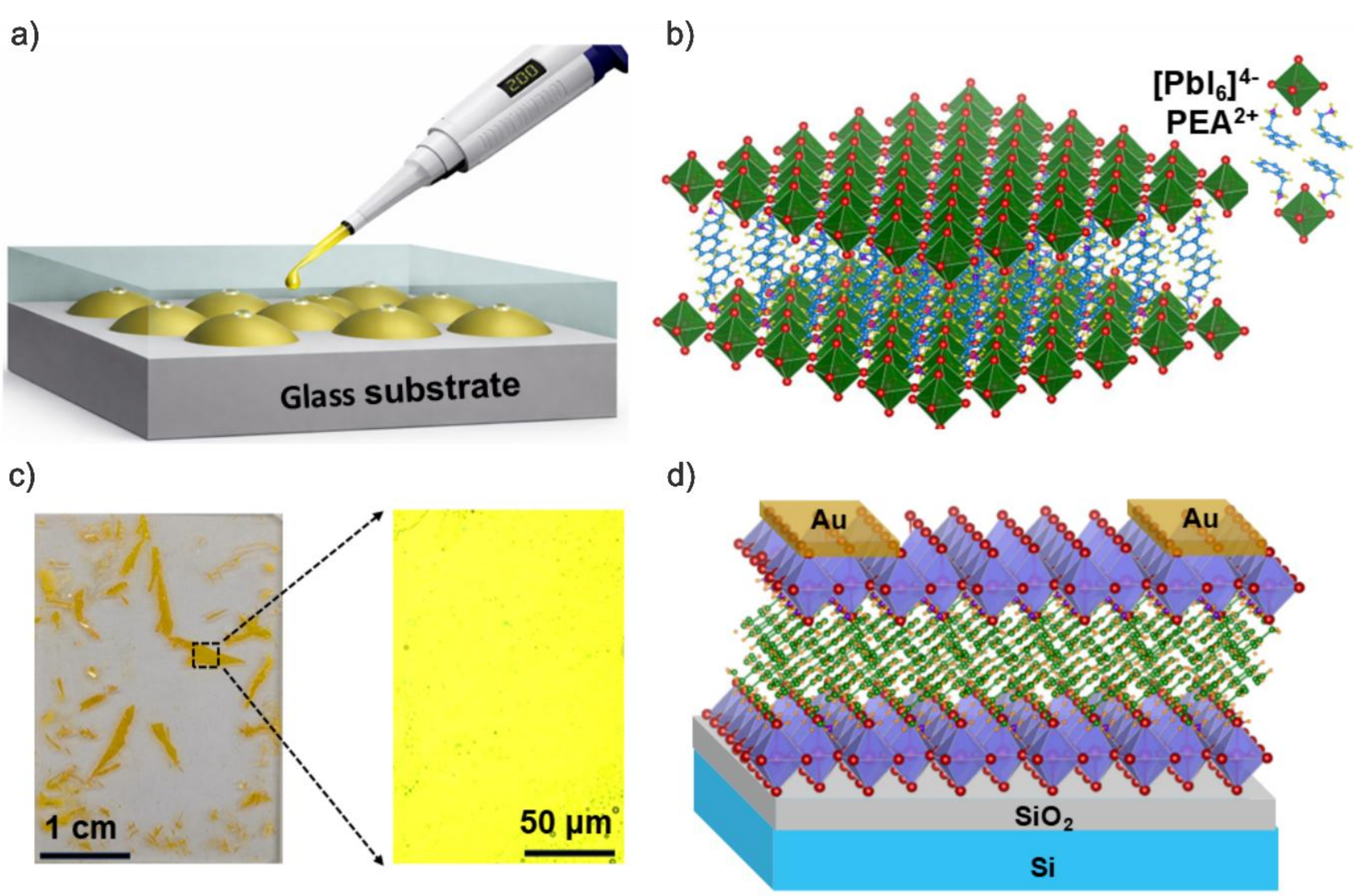

***Figure 1. Space-confined crystal growth and device schematic of 2D halide perovskites.*** *(a) Schematic illustration of the substrate stacking used for space-confined crystal growth. (b) Crystal structure of $(PEA)_2PbI_4$ 2D halide perovskite. (c) Photograph of an as-grown sample on a glass substrate; insets show optical microscopy images recorded at 20× magnification. (d) Schematic of the device architecture based on as-grown 2D halide perovskite single crystals on a $Si/SiO_2$ substrate.*

We first examine the structural and optoelectronic properties of the two-dimensional (2D) Ruddlesden–Popper perovskites $(PEA)_2PbI_4$ (phenethylammonium $C_6H_5CH_2CH_2NH_3^+$, abbreviated as PEA) and its fluorinated analogue $(PEA\text{-}F)_2PbI_4$, which together provide a well-defined materials platform for probing lattice–carrier interactions. High-quality single crystals were grown using a space-confined antisolvent vapor-assisted crystallization method (Fig. 1a) [15], yielding large-area, plate-like crystals with well-defined morphology (Fig. 1c) (details of the synthesis procedure can be found in the Methods section and the Supplementary Information (figure S1). Fig. 1c presents an optical image of the as-grown sample, showing large needle-like single crystals with centimetre-scale lengths and widths spanning from tens to several hundreds of micrometres. Following growth, the substrates were gently separated using a blade, and a representative single crystal was selected for subsequent device fabrication. Fig. 1d illustrates a schematic representation of the device architecture based on as-grown 2D halide perovskite single crystals integrated on a $Si/SiO_2$ substrate (details of the device fabrication can be found in the methods section). The crystal structure of $(PEA)_2PbI_4$ consists of single inorganic layers of corner-sharing $PbI_6$ octahedra separated by bilayers of phenethylammonium ($PEA^+$) cations (Fig. 1b). This layered architecture forms a natural quantum-well, in which the inorganic sheets act as semiconducting channels while the organic spacers behave as an insulator.

X-ray diffraction (XRD) measurements (Fig. S2a and Fig. S2b) reveal a series of sharp (00l) reflections for both $(PEA)_2PbI_4$ and $(PEA\text{-}F)_2PbI_4$ thin films, indicating a high degree of crystallographic orientation (details on thin film fabrication can be found in the methods section). Both films exhibit a series of intense, regularly spaced diffraction peaks at 5.4°, 10.8°, 16.2°, and higher orders, which can be indexed to the (001), (002), (003), …, (00*l*) reflections of layered $(PEA)_2PbI_4$ as well as for $(PEA\text{-}F)_2PbI_4$ [31]. Compared to $(PEA)_2PbI_4$, the diffraction peaks of $(PEA\text{-}F)_2PbI_4$ show a slight shift toward smaller angles, corresponding to an increased interlayer spacing, as highlighted in the enlarged view in Figure S2b (supporting information). Optical microscope images of the $(PEA)_2PbI_4$ and $(PEA\text{-}F)_2PbI_4$ thin films are shown in Figure S3a and Figure S3b, respectively (see Supplementary Information). Fig.S4a and Fig. S4b presents the room-temperature photoluminescence (PL) spectra and optical absorption recorded under 395 nm excitation. Narrow photoluminescence (PL) emission peaks are observed at ~521 nm for $(PEA)_2PbI_4$ and ~525 nm for $(PEA\text{-}F)_2PbI_4$, accompanied by pronounced excitonic absorption features [32, 33].

## Calculations of the phonon modes

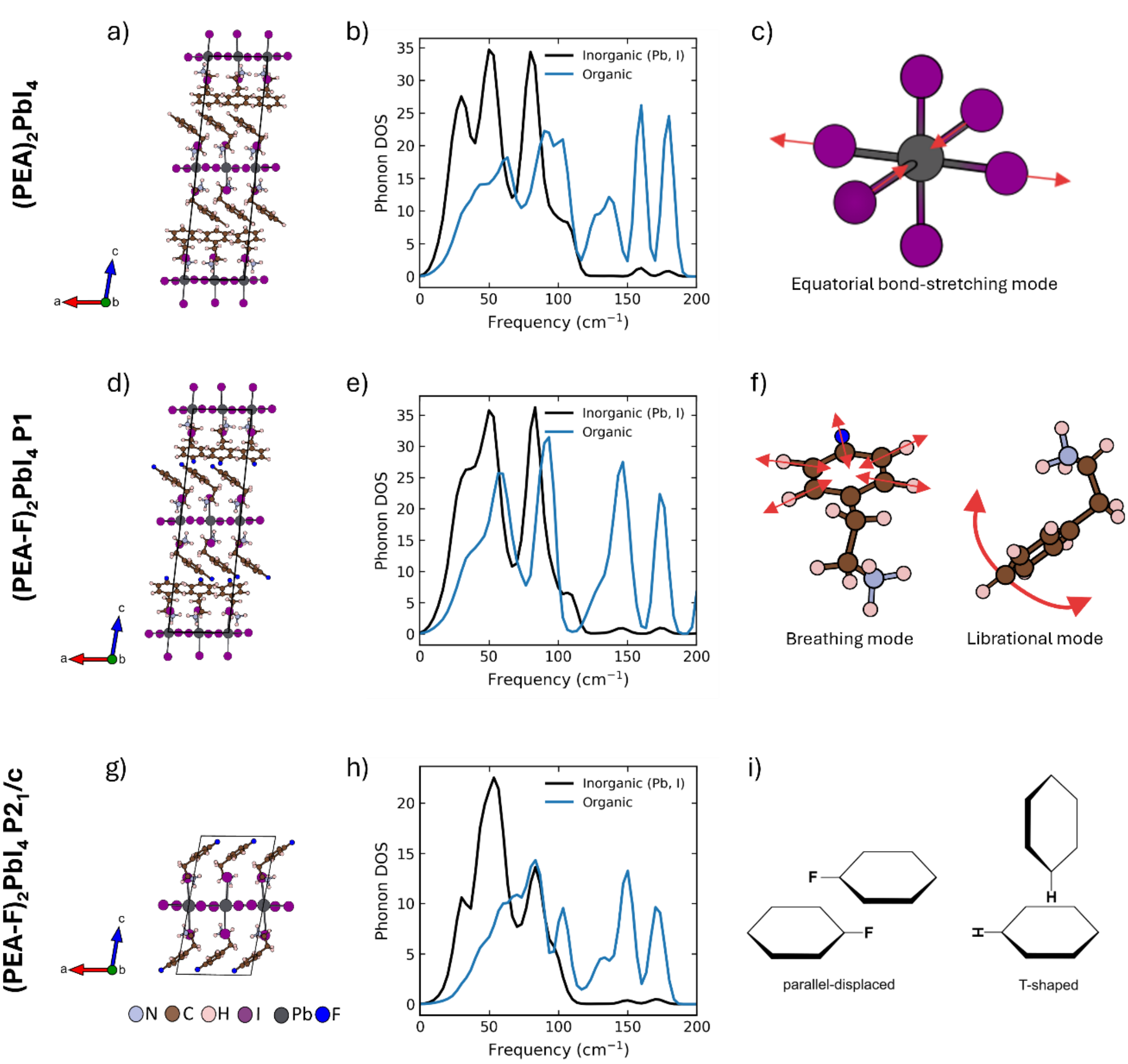

*Figure 2. Schematic representation of the studied structures and their vibrational properties*

*Schematic overview of the investigated structures. (a) $(PEA)_2PbI_4$, (d) $(PEA\text{-}F)_2PbI_4$ (P1), and (g) $(PEA\text{-}F)_2PbI_4$ ($P2_1/c$), where P1 and $P2_1/c$ denote the corresponding space groups. The corresponding projected phonon density of states (PDOS), highlighting contributions from the inorganic Pb–I octahedra and the organic cations, are shown in (b), (e), and (h), respectively. (c) Symmetric Pb–I equatorial bond-stretching mode observed in $(PEA)_2PbI_4$ and (PEA-*

*$F)_2PbI_4$ (P1) around 100 $cm^{-1}$. (f) Representative cation modes, including the phenyl rotational/PEA liberational mode in (PEA-F)$_2$PbI$_4$ ($P2_1/c$) and the PEA liberational mode in (PEA-F)$_2$PbI$_4$ (P1) and (PEA)$_2$PbI$_4$. (i) Comparison of packing configurations of the $(PEA)^+$ and $(PEA\text{-}F)^+$ cations.*

Multiple crystal structures of (PEA-F)$_2$PbI$_4$ and (PEA)$_2$PbI$_4$ have been reported in the literature. To identify suitable starting configurations, we assess their thermodynamic stability. To sample a broader range of cation packing motifs, literature-reported structures [34–36], along with additional configurations generated via substitution, were relaxed using density functional theory (DFT). For (PEA-F)$_2$PbI$_4$, the (PEA)$_2$PbI$_4$ structure reported by Du et al. [34] was modified by replacing the hydrogen atom at the para position of the phenyl ring with fluorine. Conversely, a (PEA)$_2$PbI$_4$ structure was generated from the (PEA-F)$_2$PbI$_4$ configuration of Slavney et al. [35] by substituting fluorine with hydrogen. Computational details, as well as optimized structures and energies, are provided in the Supporting Information (SI) (Table S1).

Because the systems differ in the number of atoms, total energies were normalized per Pb atom. The experimentally reported structures of Du et al. [34] (Fig. 2a) and Slavney et al. [35] (Fig. 2g) are identified as the most stable configurations. Notably, these structures exhibit distinct crystal symmetries and markedly different lattice parameters along the stacking direction (c). (PEA)$_2$PbI$_4$ adopts P1 symmetry with a large lattice parameter (c = 32.555 Å; Fig. 2a) consisting two cation layers, whereas the fluorinated system crystallizes in the $P2_1/c$ space group (c = 16.443 Å; Fig. 2g) with a single cation layer. As discussed below, the differences in the phonon spectra of these compounds arise from the chemical nature of the cation, as well as variations in packing density and crystal structure. To disentangle these effects, we compare three representative structures (Fig. 2): (PEA)$_2$PbI$_4$ (P1), (PEA-F)$_2$PbI$_4$ (P1), and (PEA-F)$_2$PbI$_4$ ($P2_1/c$), where P1 and $P2_1/c$ denote the corresponding space groups.

**Effect of fluorination on phonon properties**

We first compare the phonon properties of $(PEA)_2PbI_4$, reported by Du et al. [34] (Fig. 2a), with the corresponding $(PEA\text{-}F)_2PbI_4$ structure generated via substitution (Fig. 2d). The projected phonon density of states (PDOS), resolved into inorganic Pb–I and organic contributions, is shown in Figs. 2b and 2e. The primary differences occur in the 90–175 $cm^{-1}$ range. We focus on the range around 100 $cm^{-1}$, as it plays a major role in the experiments to be discussed below. $(PEA)_2PbI_4$ exhibits a double peak in the PDOS (~90–110 $cm^{-1}$) connected to vibrations of the organic molecule, whereas $(PEA\text{-}F)_2PbI_4$ shows a single peak just below 100 $cm^{-1}$. This shift is consistent with the increased mass of the $(PEA\text{-}F)^+$ cation. Within the harmonic approximation ($\omega \propto \sqrt{(1/m)}$), the mass ratio (~1.147) yields a scaling factor of ~0.93, in good agreement with the observed downshift of molecular modes. Mode analysis supports this interpretation: the ~100 $cm^{-1}$ mode in $(PEA\text{-}F)_2PbI_4$ corresponds to the ~110 $cm^{-1}$ mode in $(PEA)_2PbI_4$, while modes near 80 $cm^{-1}$ remain largely unchanged due to their dominant iodine character. Consequently, the organic PDOS peak narrows upon fluorination. Modes in this range also involve symmetric Pb–I equatorial stretching (Fig. 2c), indicating coupling between organic and inorganic vibrations.

**Impact of ligand packing**

We next examine the role of molecular packing by comparing two $(PEA\text{-}F)_2PbI_4$ structures: the substituted bilayer structure (Fig. 2d) and the experimentally reported monolayer structure by Slavney et al. [35] (Fig. 2g). In the bilayer configuration, phenyl rings adopt a T-shaped $\pi$–$\pi$ stacking, whereas the monolayer exhibits a parallel-displaced arrangement (Fig. 2i). The PDOS (Figs. 2e and 2h) reveals pronounced differences near 100 $cm^{-1}$. In the bilayer structure, these modes correspond to a liberational motion of the organic cations as a whole. In contrast, the monolayer exhibits a liberational motion of the molecules that is driven by a rotation of the phenyl ring around the bond to the ethyl group [37], where the phenyl rings of the stacked PEA molecules rotate in phase. Notably, Pb–I equatorial bond-stretching contributions are smaller in this region for the monolayer, are accompanied by an out-of-plane Pb-I bond-tilting mode (see movies of the vibration modes of both systems in supplementary video VS1-VS4). In summary, while fluorination induces predictable mass-driven frequency shifts, the dominant factor governing the phonon response around 100 $cm^{-1}$ is molecular packing. This determines whether vibrational modes are dominantly liberational or involve phenyl ring rotations, thereby fundamentally reshaping the vibrational landscape around 100 $cm^{-1}$.

**Bias-dependent Raman response of 2D halide perovskite thin films**

Raman spectroscopy, a powerful tool to probe the phonons in solids, has been widely employed as a non-destructive technique to elucidate the electronic and vibrational properties of two-dimensional materials [38, 39], including the structural and lattice dynamics of halide perovskites [40, 41]. Here, we directly probe octahedral vibrations in halide perovskites by tracking their phonon modes under an in-plane electric field using room-temperature Raman spectroscopy. This approach enables detection and analysis of field-induced modifications in electron–phonon coupling within the octahedral framework.

The Raman spectra of $(PEA)_2PbI_4$ thin films (channel length: 10 μm) recorded at room temperature under unbiased and 10 V biased condition (corresponding electric field, E= $10^6$ V/m) shown in Fig. 3a. The spectra were background-subtracted and fitted using Lorentzian functions, resolving five distinct Raman modes labelled 1–5 (Figure S6a) (detailed fitting parameters and experimental procedures are provided in the SI Figure S7-S8). For pristine 2D $(PEA)_2PbI_4$ thin films, Peak 1 at ~55 $cm^{-1}$ is primarily attributed to Pb–I bond bending vibrations [27, 42]. Peaks 2 and 3, located at 72 and 82 $cm^{-1}$, are assigned to I–Pb–I bending modes [27, 43]. Peak 4, centred at ~98 $cm^{-1}$, corresponds to the symmetric stretching of the Pb–I bond, while Peak 5 around 134 $cm^{-1}$ is assigned to the bending and stretching vibrations of the large organic cations [27, 44]. The Raman response exhibits quantitatively significant changes under a 10 V electrical bias. To rigorously assess these variations, we performed systematic peak fitting using Lorentzian functions and extracted the corresponding full width at half maximum (FWHM) values (see table S11 in SI). Notably, a pronounced and selective broadening is observed for the mode near ~98 $cm^{-1}$, which is primarily associated with the equatorial Pb–I stretching vibration (Fig. 3b). The FWHM of this mode increases from 12.9 $cm^{-1}$ under zero bias to 15.2 $cm^{-1}$ under applied bias, corresponding to a broadening of approximately 2.3 $cm^{-1}$. Given the spectral resolution of the Raman system (~1 $cm^{-1}$), this change significantly exceeds the instrumental limit and therefore reflects an intrinsic lattice response to the applied electric field. In contrast, the remaining modes exhibit negligible variation, highlighting the mode-selective nature of the field-induced lattice dynamics.

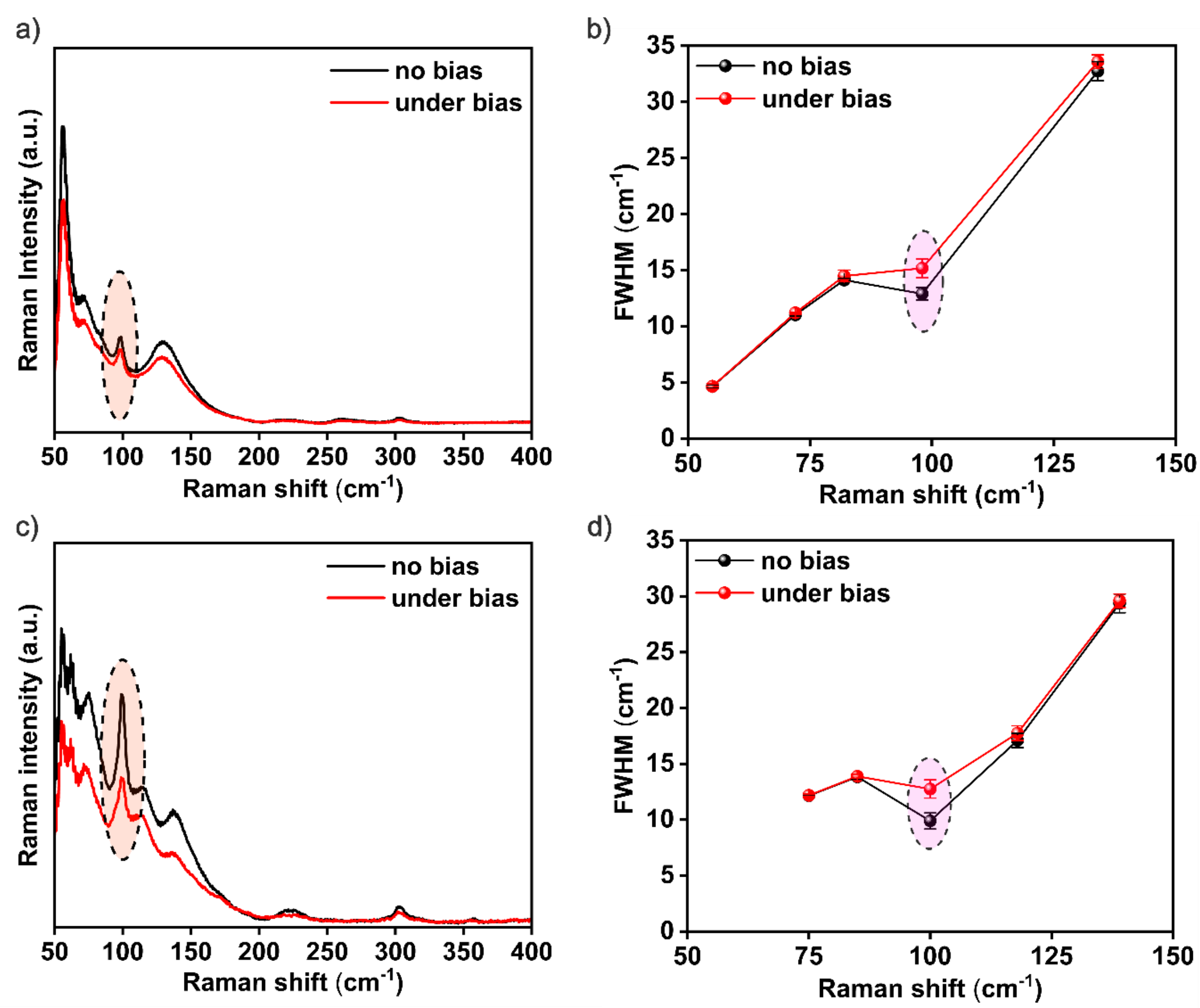

***Figure 3. Bias-dependent Raman response of 2D halide perovskite thin films.*** *(a) Raman spectra of (PEA)$_2$PbI$_4$ (n = 1) thin films deposited on Si/SiO$_2$ substrates, measured at 300 K under unbiased and biased conditions using 633 nm laser excitation after baseline correction. (b) Corresponding full width at half maximum (FWHM) of the Raman modes of (PEA)$_2$PbI$_4$ (n = 1) under unbiased and biased conditions. (c) Raman spectra of (PEA-F)$_2$PbI$_4$ (n = 1) thin films on Si/SiO$_2$ substrates, recorded at 300 K under unbiased and biased conditions with 633 nm excitation after baseline correction. (d) Corresponding full width at half maximum (FWHM) of the Raman modes of (PEA-F)$_2$PbI$_4$ (n = 1) under unbiased and biased conditions.*

An important question is whether the observed behavior is intrinsic to (PEA)$_2$PbI$_4$ or represents a more general characteristic of two-dimensional halide perovskites. To address this, we investigated the fluorinated analogue (PEA-F)$_2$PbI$_4$ (Figure S6b in SI). We observed systematic frequency shifts and, in some cases, peak splitting, pointing to subtle but important modifications of the lattice dynamics [9]. These differences cannot be fully explained by a simple stiffening of the inorganic framework [45, 46]. Theoretical phonon density-of-states calculations reveal that the largest vibrational changes occur in the 90–175 cm$^{-1}$ region, consistent with experiment. In particular, the broad ~98 cm$^{-1}$ feature in (PEA)$_2$PbI$_4$ shifts and narrows upon fluorination due to both the increased mass of the fluorinated cation and, more importantly, changes in molecular packing. Mode analysis shows that vibrations near ~100 cm$^{-1}$ are hybrid organic–inorganic modes involving coupled organic cation motion and symmetric Pb–I equatorial stretching. In the fluorinated bilayer structure, T-shaped π–π interactions promote coupled librational modes with significant Pb–I character, whereas alternative packing motifs favor more localized phenyl-ring rotational motion with reduced Pb–I participation. Consequently, fluorination redistributes the vibrational manifold near ~100 $cm^{-1}$ in both frequency and character, explaining the experimentally observed Raman shifts, peak splitting.

Fig. 3c presents the Raman spectra of the $(PEA\text{-}F)_2PbI_4$ thin-film device (channel length: 20 μm) recorded under no bias and under an applied bias of 20 V (corresponding electric field, E= $1\times10^6$ V/m). In close analogy to the $(PEA)_2PbI_4$ sample, spectral variations in $(PEA\text{-}F)_2PbI_4$ were quantitatively analysed by fitting the experimental data with a Lorentzian function (detailed fitting parameters and experimental procedures are provided in the SI Figure S9-S10). A noticeable bias dependent modification is observed for the vibrational mode centred near 100 $cm^{-1}$ (Fig. 3d), which is primarily assigned to the symmetric stretching of the Pb–I bond within the inorganic octahedral framework. Under no bias condition, this mode exhibits a FWHM of 9.89 $cm^{-1}$. Upon application of a 20 V bias, the FWHM increases distinctly to 12.75 $cm^{-1}$, corresponding to a broadening of ~2.86 $cm^{-1}$. In contrast, no appreciable variation in FWHM is detected for the remaining Raman-active modes (Fig. 3d), underscoring the selective sensitivity of the Pb–I stretching vibration to the applied electric field. The observed broadening of the 100 $cm^{-1}$ mode under bias likely signifies enhanced lattice disorder or dynamic strain within the 2D perovskite framework. Importantly, thermal effects can be excluded as the primary origin of the observed behavior: Joule heating is negligible due to the extremely low current levels (~$10^{-9}$ A), and thermal effects would produce uniform linewidth broadening across all phonon modes, which is not observed. Also, thermal effects arising from laser heating are expected to be minimal because the excitation photon energy (1.96 eV, 633 nm) lies below the optical bandgap (~2.4 eV), substantially suppressing direct interband absorption. The selective modification of the Pb–I stretching mode thus supports a carrier-mediated origin of the observed spectral changes. This provides direct spectroscopic evidence of bias dependent electron–phonon coupling in 2D halide perovskites.

A plausible interpretation of the bias-induced broadening of the Raman linewidth at ~98 $cm^{-1}$ in $(PEA)_2PbI_4$ and ~100 $cm^{-1}$ in $(PEA\text{-}F)_2PbI_4$ is that it reflects enhanced lattice polarization and dynamic strain arising from charge injection. Under an applied electric field, this mode exhibits a pronounced linewidth broadening, which can be attributed to enhanced electron–phonon coupling. Microscopically, the injected charge carriers polarize the lattice and dynamically distort the Pb–I octahedral network, thereby modulating the coupled organic–inorganic vibration. Since the electronic structure is primarily derived from Pb-6p and I-5p orbitals, such distortions directly reshape the energy landscape for charge transport. In RP perovskites, charge transport is strongly anisotropic, occurring predominantly within the inorganic sheets, while the organic spacers impose dielectric confinement. As a result, carrier motion is inherently coupled to lattice distortions within the Pb–I framework, particularly to vibrational modes that involve strong polarization of the lattice. The absence of any appreciable linewidth changes for modes below 90 $cm^{-1}$ (Pb–I bending) and above 120 $cm^{-1}$ (predominantly organic vibrations) highlights the mode-selective nature of this interaction. This selectivity indicates that only specific lattice vibrations—namely those that effectively couple electronic and polar lattice degrees of freedom participate in the field-induced response. Within this framework, the hybrid mode near ~98/100 $cm^{-1}$ strongly coupled to longitudinal optical (LO)-like polar vibrations of the Pb–I network, which mediate Fröhlich-type electron–phonon coupling [47, 48]. In this mechanism, charge carriers interact with long-range electric fields generated by polar lattice oscillations, leading to polaron formation and enhanced phonon scattering. In $(PEA\text{-}F)_2PbI_4$, the electron-withdrawing nature of fluorine modifies the ligand packing and enhances dielectric confinement, thereby strengthening the coupling between carriers and polar lattice modes. This is consistent with the experimentally observed increase in the inorganic-to-organic Raman intensity ratio (I_inorganic/I_organic), observed for $(PEA\text{-}F)_2PbI_4$ compared to $(PEA)_2PbI_4$ (see Tables S3-S6 in SI) as well as the more pronounced linewidth modulation under bias. The fluorinated system thus exhibits stronger carrier–lattice interaction, reflecting an enhancement of Fröhlich-type coupling mediated by the modified dielectric environment.

**Bias-dependent Raman response of 2D halide perovskite single crystals**

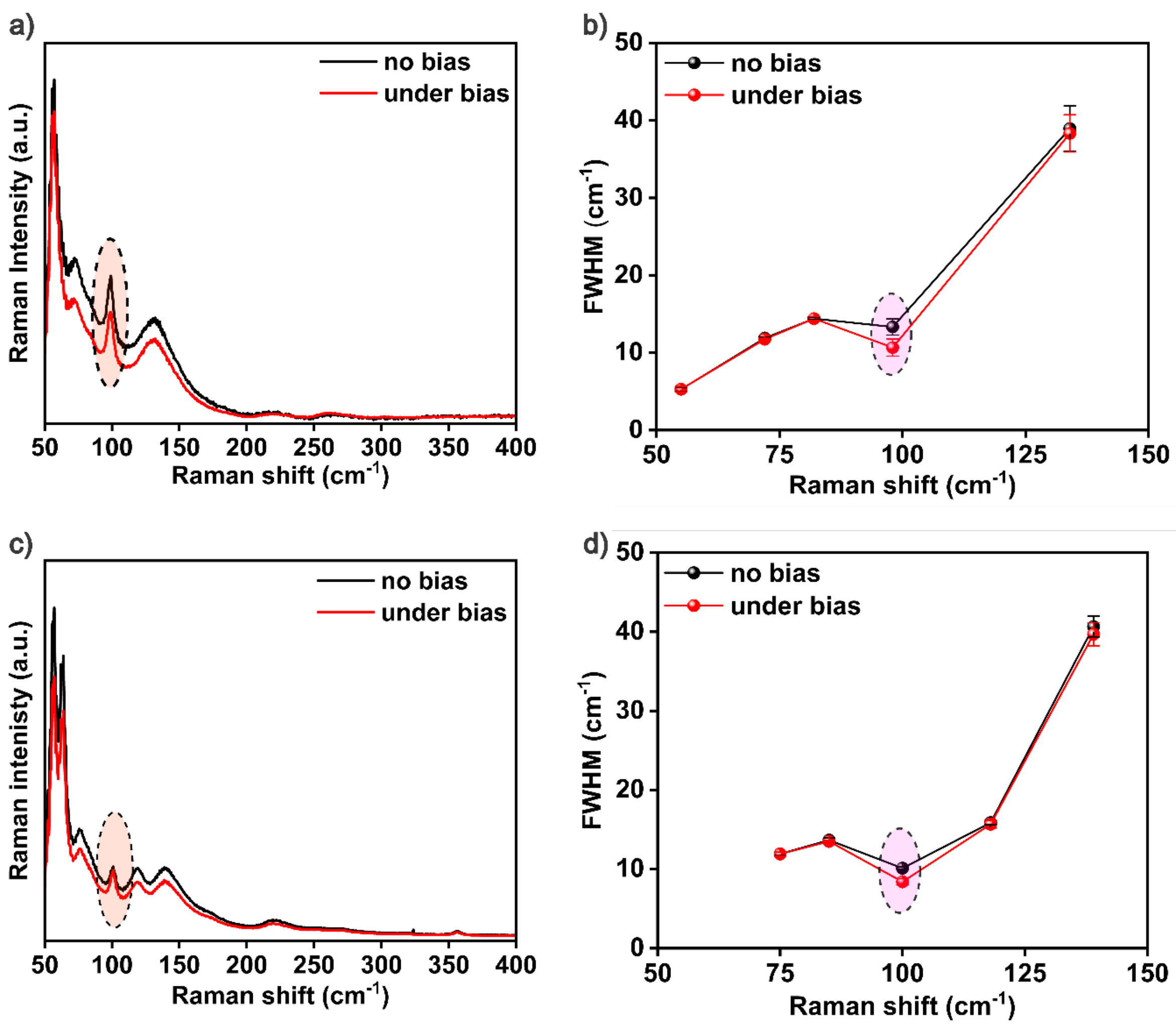

***Figure 4. Bias-dependent Raman response and charge transport in 2D halide perovskite single crystals.***

*(a) Raman spectra of $(PEA)_2PbI_4$ (n = 1) single crystals grown on $Si/SiO_2$ substrates, measured at 300 K under unbiased and biased conditions using a 633 nm laser excitation. (b) Corresponding full width at half maximum (FWHM) of the Raman modes of $(PEA)_2PbI_4$ (n = 1) under unbiased and biased conditions. (c) Raman spectra of $(PEA\text{-}F)_2PbI_4$ (n = 1) single crystals on $Si/SiO_2$ substrates, recorded at 300 K under unbiased and biased conditions with a*

*633 nm excitation. (d) Corresponding FWHM of the Raman modes of $(PEA\text{-}F)_2PbI_4$ (n = 1) under unbiased and biased conditions.*

Thin films provide a device-relevant platform, as most practical optoelectronic devices are based on thin-films. In contrast, single crystals, with their high crystallinity and low defect density, are ideal for probing intrinsic lattice responses and quantifying subtle electric-field-induced electron–phonon coupling. Studying both systems together links intrinsic phonon behavior to practical device performance. The Raman spectra of $(PEA)_2PbI_4$ single crystals recorded at room temperature under both unbiased and biased conditions shown in Fig. 4a. The Raman modes of the single crystals closely mirror those observed in the thin films. To probe bias dependent variations, an 80 V in-plane bias was applied across the crystal (channel length ~ 70 µm, corresponding electric field E= $1.14\times10^6$ V/m) and the Raman response was monitored (Fig. 4a). Quantitative Lorentzian fitting reveals a pronounced narrowing of the 98 $cm^{-1}$ mode. Its FWHM decreases from 13.33 $cm^{-1}$ at no bias condition to 10.67 $cm^{-1}$ under bias condition, corresponding to a linewidth narrowing of ~2.66 $cm^{-1}$ well above the instrumental resolution (~1 $cm^{-1}$) (detailed fitting parameters and experimental procedures are provided in the SI Figure S11-S12). In contrast, the other modes show negligible changes in linewidth (Fig. 4b). Similarly, for $(PEA\text{-}F)_2PbI_4$ single crystals, the Raman spectra recorded at room temperature under both zero-bias and biased conditions (channel length ~ 70 µm, applied voltage 100 V, corresponding electric field E= $1.42\times10^6$ V/m) are shown in Fig. 4c. The spectral positions and assignments closely match those of the corresponding thin films, but the single crystals exhibit sharper features, reflecting their superior crystallinity and reduced structural disorder. The narrower linewidths indicate diminished phonon scattering from defects and grain boundaries, allowing a clearer resolution of the intrinsic lattice vibrations. Lorentzian fitting reveals a selective narrowing of the 100 $cm^{-1}$ mode, corresponding to the symmetric equatorial Pb–I stretching vibration, with its FWHM decreasing from 10.12 $cm^{-1}$ at zero bias to 8.4 cm$^{-1}$ under applied voltage a reduction of ~1.72 cm$^{-1}$ (detailed fitting parameters and experimental procedures are provided in the SI Figure S13-S14). The linewidths of the other modes remain largely unchanged (Fig. 4d), indicating that the effect is localized to the inorganic framework. Within the $PbI_6$ octahedra, the equatorial Pb–I stretching mode corresponds to a polar longitudinal optical (LO) phonon, which is expected to strongly couple to charge carriers via the Fröhlich interaction. The strength of this interaction is estimated using the conventional Fröhlich coupling constant [49, 50],

$$\alpha = \frac{e^2}{4\pi\varepsilon_0\hbar}\left(\frac{1}{\varepsilon_\infty} - \frac{1}{\varepsilon_s}\right)\left(\frac{m^*}{2\hbar\omega_{LO}}\right)^{\frac{1}{2}}$$

where $e$ is the electron charge, $\varepsilon_s$ is static dielectric constant, $\varepsilon_\infty$ is high-frequency dielectric constant, $\varepsilon_0$ is the vacuum permittivity, $\hbar$ is the reduced Planck constant, $m^*$ is the carrier effective masses and $\omega_{LO}$ is the LO phonon frequency and is further enhanced in two-dimensional (2D) perovskites due to dielectric confinement. Within this framework, the applied electric field modulates the injected carrier density, thereby tuning the effective electron–phonon interaction strength. The observed bias-dependent linewidth variations therefore constitute spectroscopic signatures consistent with enhanced carrier–phonon coupling, rather than a global lattice perturbation. A comprehensive comparison of the bias-dependent evolution of the FWHM as a function of Raman shift for both $(PEA)_2PbI_4$ and $(PEA\text{-}F)_2PbI_4$ thin film and single crystal is presented in Table S11 (SI). This table summarize the extracted FWHM under both unbiased and biased conditions, enabling a direct evaluation of the electric-field-induced changes in the vibrational response of individual Raman peaks. Statistical analysis of Raman linewidths (Fig. S15) reveals narrower FWHM in single crystals than in thin films under biased condition.

## 3. Discussion

### Phonon lifetime in 2D halide perovskites

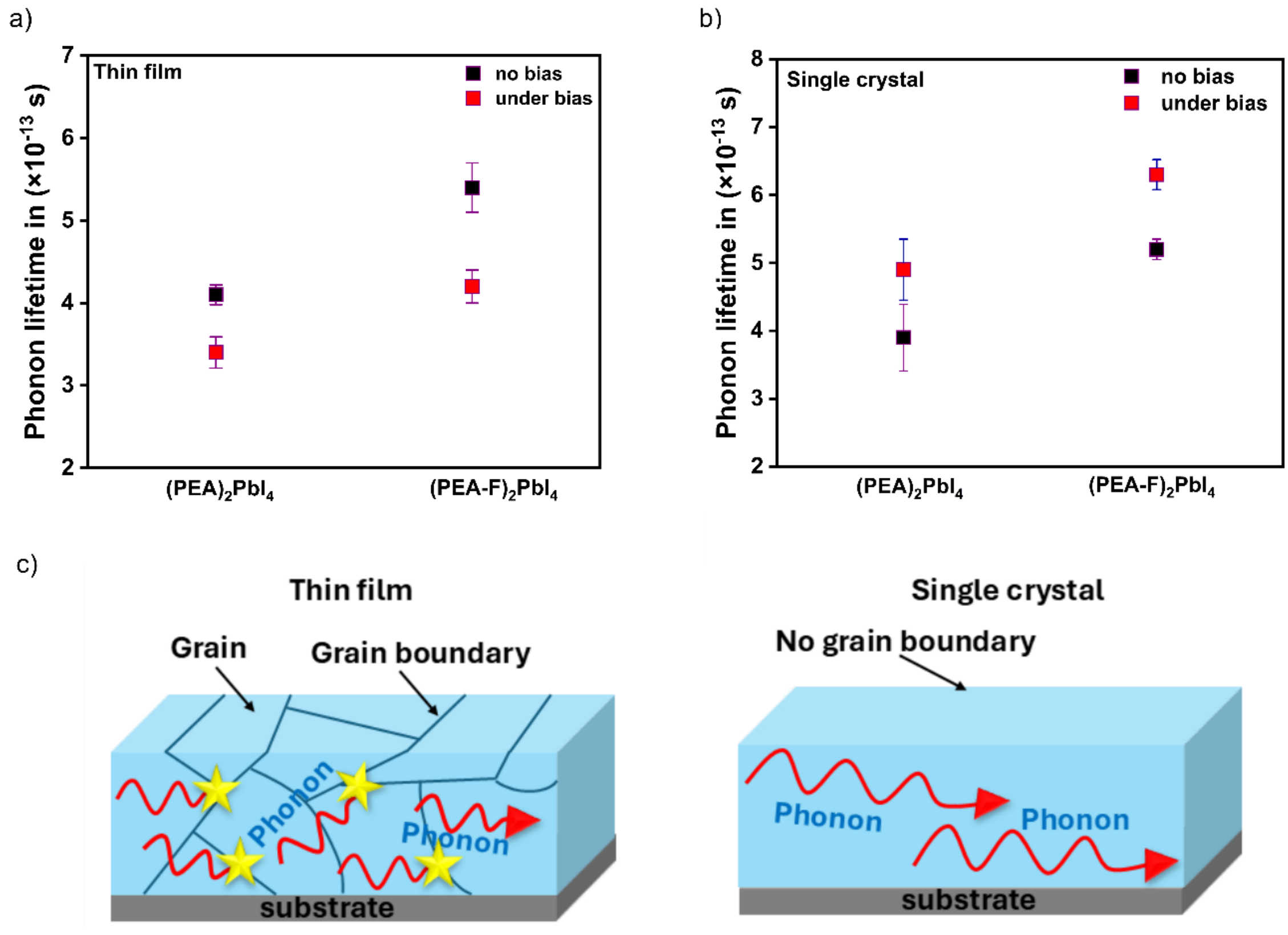

***Figure 5. Phonon lifetime in 2D halide perovskites***

*Phonon lifetimes derived from the ~98 $cm^{-1}$ Raman mode of $(PEA)_2PbI_4$ (n = 1) and ~100 $cm^{-1}$ mode of $(PEA\text{-}F)_2PbI_4$ (n = 1) at 300 K under unbiased and biased conditions for (a) thin films and (b) single crystals. Error bars represent the standard error. (c) Schematic illustration of phonon propagation pathways in thin films and single crystals.*

Phonon lifetimes play a central role in determining the physical properties of halide perovskites, as they strongly influence thermal transport, optoelectronic response, and charge-carrier dynamics. In halide perovskite, phonons typically possess very short lifetimes, often on the picosecond timescale, reflecting the pronounced lattice anharmonicity [51]. Fig. 5a presents the extracted phonon lifetimes for thin film $(PEA)_2PbI_4$ and $(PEA\text{-}F)_2PbI_4$ under both unbiased and biased conditions. The lifetimes were determined from the Raman-active modes located at ~98 $cm^{-1}$ for $(PEA)_2PbI_4$ and ~100 $cm^{-1}$ for $(PEA\text{-}F)_2PbI_4$. To estimate the phonon lifetimes from the Raman linewidths, we employed the energy–time uncertainty relation, using the measured full width at half maximum (FWHM) of the corresponding Raman peaks [52].

$$\frac{\Delta E}{\hbar} = \frac{1}{\tau}$$

where $\Delta E$ is the Raman FWHM in units of $cm^{-1}$, and $\hbar$=5.3 × $10^{-12}$ $cm^{-1}$ s. The extracted phonon lifetimes reveal a clear bias-dependent evolution that differs distinctly between thin-film and single-crystal devices. For the $(PEA)_2PbI_4$ thin film, the phonon lifetime under unbiased conditions is estimated to be 4.1 (±0.12) × $10^{-13}$ s, which decreases to 3.4 (±0.19) × $10^{-13}$ s upon the application of an electrical bias. A similar trend is observed for the $(PEA\text{-}F)_2PbI_4$ thin film, where the phonon lifetime reduces from 5.4 (±0.3) × $10^{-13}$ s in the unbiased state to 4.2 (±0.2) × $10^{-13}$ s under bias. These results indicate that the application of an electric field enhances phonon scattering processes in thin-film devices, leading to a shortening of the phonon lifetime. Such behavior can be attributed to the presence of grain boundaries and defect-mediated scattering pathways that become more active under an applied electric field. In contrast, the single-crystal devices exhibit the opposite behavior. For the $(PEA)_2PbI_4$ single crystal, the phonon lifetime increases from 3.9 (±0.49) × $10^{-13}$ s under unbiased condition to 4.9 (±0.45) × $10^{-13}$ s under electrical bias (Fig. 5b). Similarly, for the $(PEA\text{-}F)_2PbI_4$ single crystal, the phonon lifetime increases from 5.2 (±0.15) × $10^{-13}$ s to 6.3 (±0.11) × $10^{-13}$ s under bias (Fig. 5b). The observed increase in phonon lifetime suggests that, in the absence of significant structural disorder, the applied electric field may stabilize the lattice vibrations or reduce certain phonon scattering channels in the highly ordered single-crystal lattice.

Experimentally, the Raman linewidth (Γ) provides direct insight into phonon scattering processes and can be expressed as:

$$\Gamma = \Gamma_{anh} + \Gamma_{def} + \Gamma_{e-ph}$$

where $\Gamma_{anh}$ arises from intrinsic anharmonic phonon decay, $\Gamma_{def}$ from defect- and disorder-induced scattering, and $\Gamma_{e-ph}$ from electron–phonon interactions. The mode-selective evolution of Γ under bias therefore directly reflects changes in $\Gamma_{e-ph}$, providing spectroscopic signatures consistent with field-induced modulation of carrier–phonon coupling. The observed linewidth variations (~2–3 cm$^{-1}$) correspond to a substantial (~20–30%) change in phonon lifetime, highlighting the sensitivity of LO phonons to carrier injection even under modest electric fields. The contrasting bias response observed in thin films and single crystals highlights the interplay between extrinsic disorder and intrinsic lattice dynamics. In thin films, the linewidth broadening under bias indicates an increase in total phonon scattering rate, dominated by enhanced $\Gamma_{def}$ and disorder-assisted $\Gamma_{e-ph}$. Grain boundaries, defects, and local strain fields provide additional scattering pathways that couple to charge carriers under an applied field, leading to increased phonon damping. In this regime, electron–phonon interactions are strongly influenced by the defect, and the applied bias effectively amplifies disorder mediated scattering processes. In contrast, single crystals exhibit linewidth narrowing under bias, indicating a reduction in effective phonon scattering and a corresponding increase in phonon lifetime. In the absence of significant disorder ($\Gamma_{def}\rightarrow 0$), the linewidth is governed primarily by intrinsic contributions. Such behavior suggests a reduction in available phonon scattering channels and points toward the emergence of non-equilibrium phonon populations. One plausible interpretation is the onset of a hot-phonon bottleneck regime [53-55], in which the emission of LO phonons by hot carriers outpaces their anharmonic decay. A schematic diagram (Fig. 5c) shows how phonons are propagated through thin film and single crystal.

Recent perspectives have highlighted that the interpretation of hot-phonon bottlenecks in halide perovskites remains controversial, underscoring the importance of complementary approaches capable of directly probing carrier–phonon interactions under device-relevant operating conditions. While further studies are required to fully validate this mechanism [56].

Fluorination of the organic spacer fundamentally reshapes lattice dynamics and charge transport in 2D halide perovskites by reconfiguring hybrid organic–inorganic phonon modes [ 57]. A direct comparison between $(PEA)_2PbI_4$ and $(PEA\text{-}F)_2PbI_4$ shows that this chemical modification alters molecular packing and crystal symmetry, thereby modulating carrier–phonon interactions. A direct comparison between $(PEA)_2PbI_4$ and $(PEA\text{-}F)_2PbI_4$ shows that fluorination alters molecular packing and crystal symmetry, thereby modulating carrier–phonon interactions. In $(PEA)_2PbI_4$, the phenyl rings adopt a T-shaped $\pi$–$\pi$ stacking arrangement, leading predominantly to collective vibrational motion of the organic cations with relatively limited phenyl-ring distortion. In contrast, fluorination in $(PEA\text{-}F)_2PbI_4$ promotes a parallel-displaced packing configuration that enhances cooperative intermolecular interactions and enables rotational motion of the phenyl rings. These differences give rise to a markedly altered phonon density of states near ~100 $cm^{-1}$, where the vibrational response evolves from predominantly collective vibrational modes with dominant equatorial Pb–I bond-stretching character to modes involving phenyl-ring rotations and coupled Pb–I bond tilting. As a result, the lattice response under an applied electric field is strongly modified. These differences give rise to a markedly altered phonon density of states near ~100 $cm^{-1}$ and directly influence the lattice response under an applied electric field. The longer phonon lifetime in $(PEA\text{-}F)_2PbI_4$ indicates reduced intrinsic phonon scattering, while its stronger bias-dependent linewidth modulation suggests enhanced carrier-mediated electron–phonon interactions under applied field. These lattice-level effects translate directly into transport, with the fluorinated crystal displaying nearly a thirty-fold increase in electrical conductivity ($1.30 \times 10^{-9}$ S $cm^{-1}$ for $(PEA)_2PbI_4$ and 37.1 × $10^{-9}$ S $cm^{-1}$ for $(PEA\text{-}F)_2PbI_4$) (Figure S16 and table S13 in SI). Together, these results establish a unified microscopic framework in which molecular-level engineering of the organic spacer governs lattice polarization, hybrid phonon dynamics, and charge transport, providing a powerful route to tune electron–phonon coupling in layered perovskites.

**4. Conclusion**

In summary, we demonstrate that operando Raman spectroscopy combined with first-principles calculations provides direct microscopic insight into electron–phonon interactions in two-dimensional halide perovskites. Bias-dependent measurements reveal highly mode-selective linewidth broadening centered near ~100 $cm^{-1}$, identifying specific lattice vibrations that couple strongly to injected charge carriers under an applied electric field. DFT calculations show that these modes possess a hybrid organic–inorganic character involving coupled motion of the organic spacer and Pb–I framework, establishing that the lattice response cannot be understood solely in terms of inorganic octahedral vibrations. Fluorination of the organic spacer fundamentally modifies the vibrational landscape through changes in molecular packing, crystal symmetry, and organic–inorganic coupling. These structural changes reshape the phonon density of states, alter phonon relaxation pathways, and enhance carrier-mediated lattice interactions, ultimately leading to substantially improved electrical conductivity in the fluorinated system. Importantly, our results identify ligand packing as a key parameter governing hybrid phonon dynamics and Fröhlich-type electron–phonon coupling in layered perovskites. Beyond revealing the microscopic origin of carrier–lattice interactions in 2D halide perovskites, this work establishes molecular-level structural engineering as an effective strategy for tuning phonon dynamics and transport properties in low-dimensional semiconductors. These findings provide a framework for the rational design of perovskite materials for optoelectronic, spintronic, and electrically driven quantum devices where controlled electron–phonon coupling is essential.

## 5. Methods:

### 5.1 Materials and chemicals

Phenethylammonium iodide (PEAI, 99% purity, Sigma Aldrich), 4-fluoro-phenethylammonium iodide (F-PEAI, ≥99%, Greatcell Solar Materials), lead iodide ($PbI_2$, 99% purity, Sigma Aldrich), γ-butyrolactone (GBL, 99%, Sigma Aldrich), and N, N-dimethylformamide (DMF, Sigma Aldrich) were used in this experiment.

### 5.2 Thin film fabrication of 2D halide perovskite:

Fraunhofer substrates (pre-patterned with gold electrodes) were ultrasonically cleaned in acetone and isopropanol (IPA) for 10 minutes each. 1 M solutions were prepared by dissolving phenethylammonium iodide (PEAI) or 4-fluoro-phenethylammonium iodide (F-PEAI) and $PbI_2$ in DMF, followed by stirring at 80 °C for 5 hours. The cleaned substrates were placed on a spin coater, and 10 μL of the perovskite solution was deposited onto each substrate. Spin coating was performed at 2500 rpm for 30 seconds, after which the substrates were annealed on a hot plate at 70 °C for 10 minutes.

### 5.3 Synthesis and device fabrication of 2D halide perovskite single crystal:

Two-dimensional perovskite single crystals were synthesized using an antisolvent vapor–assisted crystallization (AVC) method. Initially, 1 M precursor solutions were prepared by dissolving phenethylammonium iodide (PEAI) or 4-fluorophenethylammonium iodide (F-PEAI) together with $PbI_2$ in γ-butyrolactone (GBL). The solutions were stirred at 80 °C for 4 h to ensure complete dissolution and formation of a homogeneous precursor. Subsequently, 3 μL of the precursor solution was drop-cast onto glass or $Si/SiO_2$ substrates, after which the droplet was gently covered with another identical substrate to form a confined thin solution layer. The substrate assembly was then placed inside a larger sealed Teflon vial containing a smaller vial filled with 2 mL of dichloromethane (DCM) as the antisolvent (Figure S1a, Supporting Information). The system was kept undisturbed at room temperature for 12 h, allowing the DCM vapor to gradually diffuse into the perovskite precursor solution. As the antisolvent vapor slowly penetrates the solution, the solubility of the perovskite precursors decreases, leading to controlled supersaturation and nucleation. This slow and diffusion-controlled crystallization process promotes the growth of highly ordered crystals within the narrow gap between the two substrates. As a result, high-quality, millimeter-scale yellow plate-like single crystals are obtained, with typical thicknesses ranging from a few to several tens of micrometers (Figure S1b, Supporting Information). The confined growth geometry and gradual antisolvent diffusion are crucial for achieving crystals with high structural quality and minimal defect density, making them suitable for detailed structural, optical, and electrical characterization. To make a device gold electrodes were deposited via e-beam evaporation at a pressure of $1 \times 10^{-6}$ mbar using a homemade shadow mask.

**5.4 Characterization:**

X-ray diffraction (XRD) patterns were recorded using a Bruker-AXS D8 Advance diffractometer with a Cu Kα X-ray source ($\lambda = 1.5418$ Å), operated in 2θ mode over a range of 4° to 40°. Optical absorption properties were evaluated using UV–visible absorption spectra collected with a Varian Cary 5000 spectrometer equipped with an integrating sphere at room temperature. Steady-state photoluminescence (PL) spectra were measured with a Horiba Jobin Yvon Fluorolog-3 fluorescence spectrometer. Current–voltage (IV) characteristics of the devices were measured using a Keithley 4200 semiconductor characterization system.

### 5.5 Raman spectroscopy

Raman spectroscopy was conducted using a linearly polarized helium–neon laser (Melles-Griot 25-LHP-928-249, $\lambda_{ex}$ = 633 nm, TEM 00, 12 MHz linewidth) with a 10 mW power before the entrance pupil of the confocal objective. The laser was focused onto the sample using a 100× objective (NA = 0.7), creating a spot size of approximately 1 $\mu m^2$. Backscattered light was collected confocally and analyzed with a Tokyo Instruments spectrometer (500 mm focal length) equipped with an 1800 lines/mm grating. Spectra were acquired over 3 minutes, yielding a spectral resolution of ~1 $cm^{-1}$. Detection was carried out with a thermoelectrically cooled CCD (Andor iDUS 420 BUV) operating at −70°C to minimize dark noise. In this study, a helium–neon laser ($\lambda$ = 633 nm; photon energy ≈ 1.96 eV) was employed for excitation. Given that the optical bandgap of the 2D halide perovskite is ≈ 2.4 eV, this wavelength was selected to be sufficiently short to enhance Raman scattering and improve the signal-to-noise ratio, while remaining long enough to avoid inducing photoluminescence from the sample.

### 5.6 DFT calculations

The DFT calculations were performed using the projector-augmented wave (PAW) method, as implemented in the Vienna Ab-initio Simulation Package [58-60]. Structural optimization was performed using the Perdew, Burke, and Ernzerhof (PBE) [61] exchange-correlation functional with D3(BJ) van der Waals corrections [62]. The energy and force convergence criteria were set to $10^{-5}$ eV and 0.02 eV/Å, respectively. The plane-wave energy cutoff was set to 500 eV. Brillouin zone sampling was performed using a 2 × 2 × 1 Gamma-centered k-point grid [63]. The resulting total energies are compared in Table SI 1.

### 5.7 Phonon Calculations with Machine Learning Force Fields:

Harmonic phonon calculations were performed using the finite-displacement supercell approach described by Togo *et al.* and implemented in Phonopy [64, 65]. The phonon density of states (pDOS) was calculated using a supercell of 3×3×1 on a 21×21×7 Γ-centered q-mesh. The smearing width of $\sigma = 0.1$ was used, as implemented in the Phonopy software.

Within this approach, forces on all atoms in the supercell are required to construct the force constants. For this, single-point calculations were performed. Due to the computational cost of DFT for such large systems, single-point force calculations were performed using machine-learned force fields (MLFFs). The MLFFs were trained following the protocol of Pols *et al.* [66], using the same model parameters. The training set was generated via on-the-fly sampling [67] as implemented in VASP [58-60], based on *ab initio* molecular dynamics (MD) simulations in the NpT ensemble with a Langevin thermostat and barostat, with a friction coefficient $\gamma = 0.5$ ps$^{-1}$. The DFT calculations employed an energy cutoff of 500 eV, an energy convergence criterion of $10^{-5}$ eV and a 2 × 2 × 1 k-point grid [63]. The SCAN [68] exchange-correlation functional was used. Details on the number of training structures and basis functions per species are provided in Table 2.

**Supplementary Information**

Supplementary Information provides detailed experimental and computational procedures and additional characterization supporting the main findings. This includes the synthesis and structural and optical characterization of $(PEA)_2PbI_4$ and $(PEA\text{-}F)_2PbI_4$ single crystals and thin films; density functional theory calculations and phonon calculations using machine-learning force fields; a detailed comparison of the Raman-active modes; and additional bias-dependent Raman measurements of both thin films and single crystals. The Supplementary Information also contains a detailed analysis of the bias-dependent Raman linewidths and phonon lifetimes, including comparisons of the full width at half maximum (FWHM) and extracted phonon lifetimes between the two materials and sample geometries. Additional electrical conductivity measurements are also provided.

**Conflict of Interest**

The authors have no conflict of interest.

**Acknowledgements**

One of the authors, T.P., would like to acknowledge the Fonds de Recherche du Québec – Nature et Technologies (FRQNT) for funding his PBEEE postdoctoral fellowship. E.O. wishes to thank the Natural Sciences and Engineering Research Council of Canada (NSERC) for funding his research program through an individual Discovery Grant. S.C. acknowledges project Ricerca@Cnr PHOTOCAT (CUP B93C21000060006). PRIN Project INTERFACE (2022HWWW3S) and PRIN PNRR22 DELPHI (P2022W9773). The authors thank Mike Pols for his efforts in preliminary explorations of phonon calculations.

# Supplementary Information

# Operando Raman Probing of Mode-Selective Electron–Phonon Coupling in Two-Dimensional Halide Perovskites

Tufan Paul[1], Helena Boom[2], Lilian Skokan[1], Viren Tyagi[2], Mostafa Shagar[1], Aditi Sahoo[1], Silvia Colella[3], Geert Brocks[2,4], Andreas Ruediger[1], Shuxia Tao[2], and Emanuele Orgiu [1]

*[1] Centre Énergie Matériaux Télécommunications, Institut national de la recherche scientifique, 1650 Blv. Lionel-Boulet, Varennes QC J3X 1P7, Canada.*

*[2] Department of Applied Physics and Science Education, Eindhoven University of Technology, 5600 MB Eindhoven, The Netherlands*

*[3] CNR NANOTEC, c/o Department of Chemistry, University of Bari, via Orabona 4, Bari 70125, Italy.*

*[4] Computational Chemical Physics, Faculty of Science and Technology and MESA+ Institute for Nanotechnology, University of Twente, 7500 AE Enschede, The Netherlands*

**Table of Contents**

## 1. Synthesis and characterization of 2D halide perovskite single crystal

### 1.1. Synthesis of 2D halide perovskite single crystal

Two-dimensional perovskite single crystals were synthesized via an antisolvent vapor–assisted crystallization (AVC) method. Initially, 1 M precursor solutions were prepared by dissolving phenethylammonium iodide (PEAI) or 4-fluorophenethylammonium iodide (F-PEAI) together with $PbI_2$ in γ-butyrolactone (GBL). The solutions were stirred at 80 °C for 4 h to ensure complete dissolution and the formation of a homogeneous precursor mixture. Subsequently, 3 μL of the precursor solution was drop-cast onto glass or $Si/SiO_2$ substrates. The droplet was then gently covered with a second identical substrate, forming a confined thin liquid layer. This substrate assembly was placed inside a larger sealed Teflon vial, which contained a smaller vial filled with 2 mL of dichloromethane (DCM) serving as the antisolvent (Figure S1a). A real-time photograph of the substrate setup during single crystal synthesis provides a crucial visual record of the experimental configuration at the moment when crystal growth is taking place (Figure S1b). The system was left undisturbed at room temperature for 12 h, allowing DCM vapor to gradually diffuse into the precursor solution. As the antisolvent vapor slowly permeates the solution, the solubility of the perovskite precursors decreases, inducing controlled supersaturation and subsequent nucleation. This diffusion-limited crystallization process enables the gradual growth of highly ordered crystals within the confined space between the two substrates. Consequently, high-quality, millimeter-scale, yellow plate-like single crystals are obtained, with typical thicknesses ranging from a few to several tens of micrometers (Figure S1c).

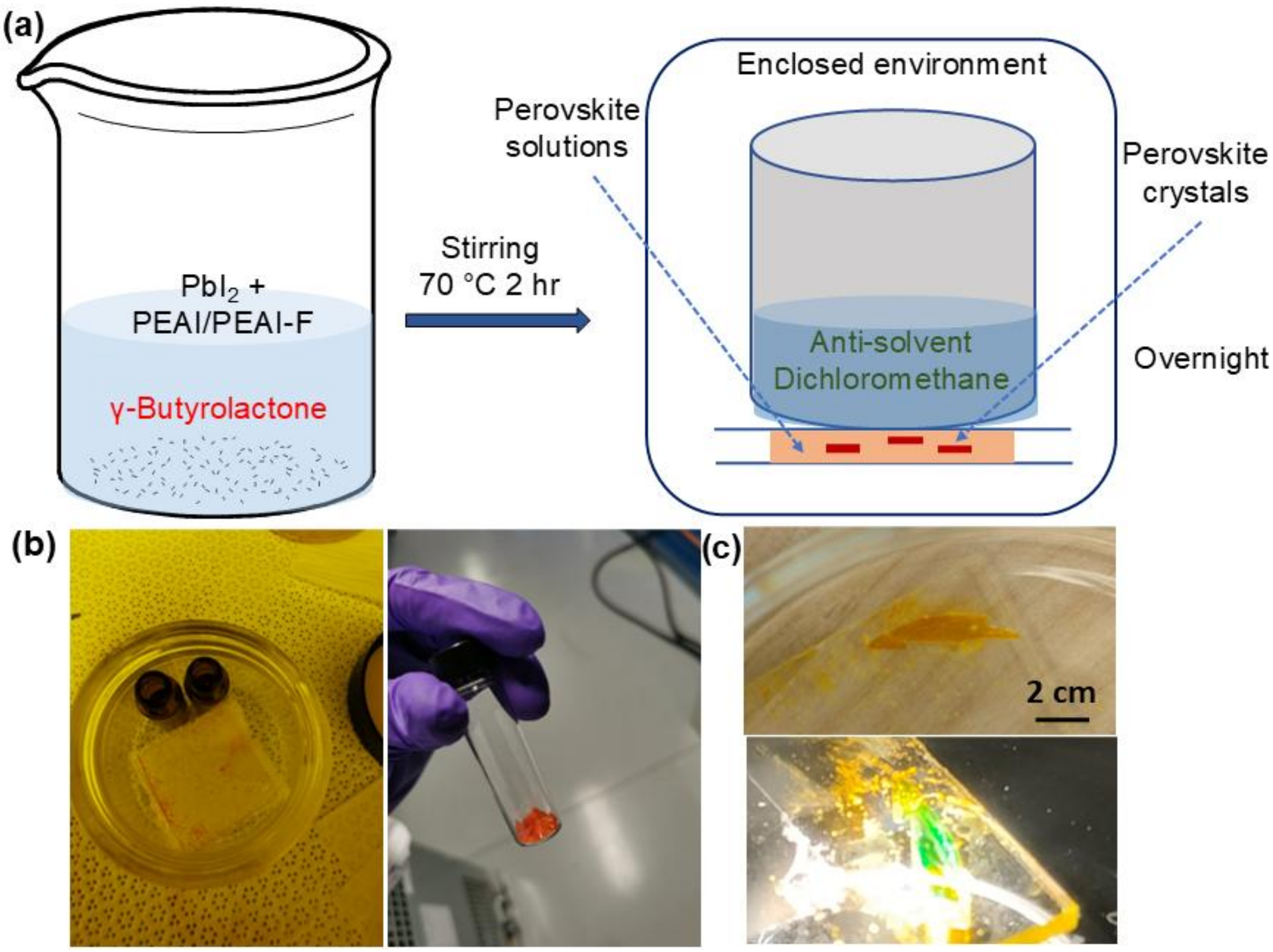

***Figure S1: Provides a comprehensive overview of the process and results of growing single crystals using the antisolvent vapor-assisted crystallization method***

*(a) Schematic illustration of the antisolvent vapor-assisted crystallization method: This panel presents a schematic diagram detailing the experimental setup and principle behind the antisolvent vapor-assisted crystallization process.* (*b) Real-Time Photograph of the Substrate Setup During Synthesis: This image shows the actual setup used during the single crystal synthesis process. The photograph captures the placement of the substrate inside the crystallization chamber, along with the reservoirs containing the precursor solution and antisolvent.* (*c) Optical Image of the As-Grown Single Crystals on a Glass Substrate: displays an optical microscope image of the single crystals formed on the glass substrate after the crystallization process. The image reveals the size, shape, and surface quality of the as-grown single crystals, demonstrating the effectiveness of the antisolvent vapor-assisted method.*

## 1.2. Structural characterization

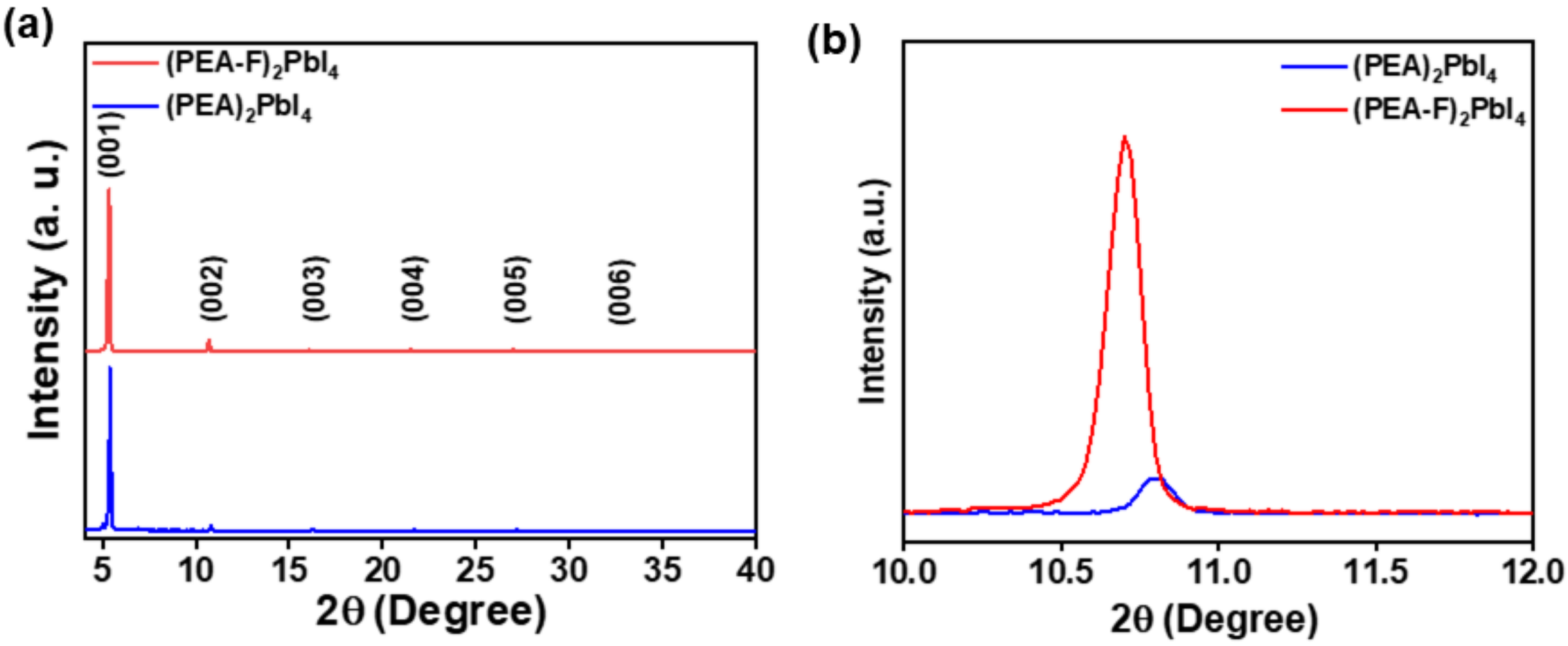

***Figure S2: Structural characterization of $(PEA)_2PbI_4$ and $(PEA\text{-}F)_2PbI_4$ thin films***

*(a) X-ray diffraction (XRD) patterns of $(PEA)_2PbI_4$ and $(PEA\text{-}F)_2PbI_4$ thin films spin-coated on glass substrates. (b) Comparison of the (002) XRD peak positions highlighting the difference between $(PEA)_2PbI_4$ and $(PEA\text{-}F)_2PbI_4$.*

### 1.3. Optical microscopy image of thin films:

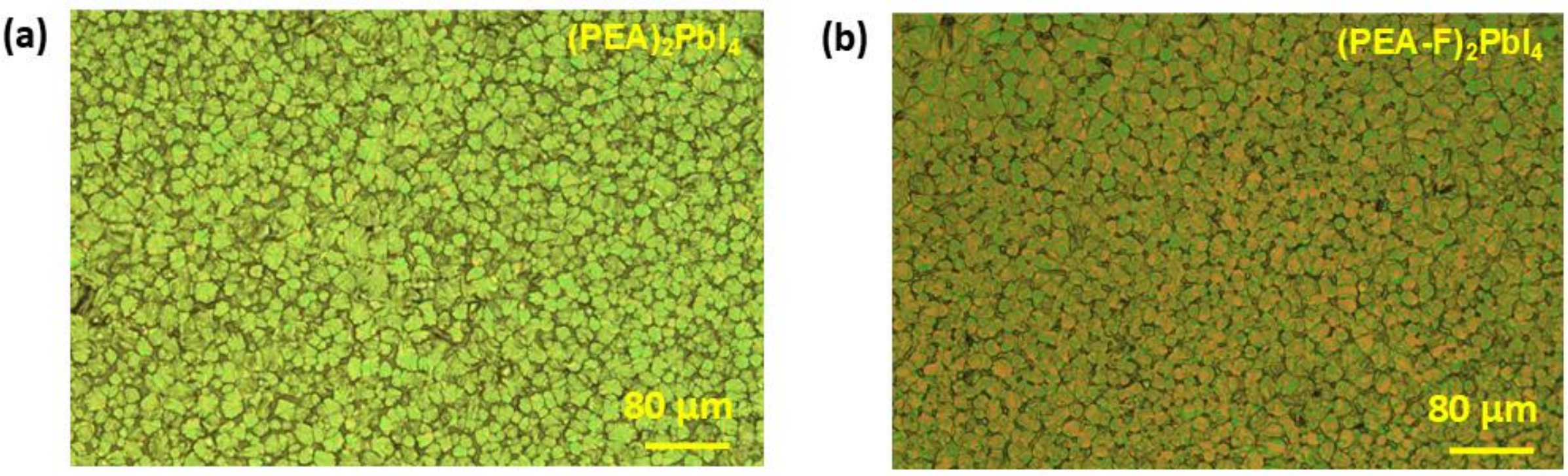

***Figure S3: Provides a comparative analysis of two types of thin films, $(PEA)_2PbI_4$ and $(PEA\text{-}F)_2PbI_4$, using optical microscopy***

*(a) Optical microscope image of $(PEA)_2PbI_4$ thin film: shows the surface morphology of a $(PEA)_2PbI_4$ thin film deposited on a glass substrate via spin coating. The optical microscope image highlights the film's uniformity, smoothness, and grain structure. (b) Optical Microscope Image of $(PEA\text{-}F)_2PbI_4$ Thin Film: this image presents the $(PEA\text{-}F)_2PbI_4$ thin film, also spin-coated onto a glass substrate. By comparing this image to the previous one, one can visually evaluate the influence of fluorine substitution (PEA-F vs. PEA) on the thin film morphology.*

## 1.4. Optical properties

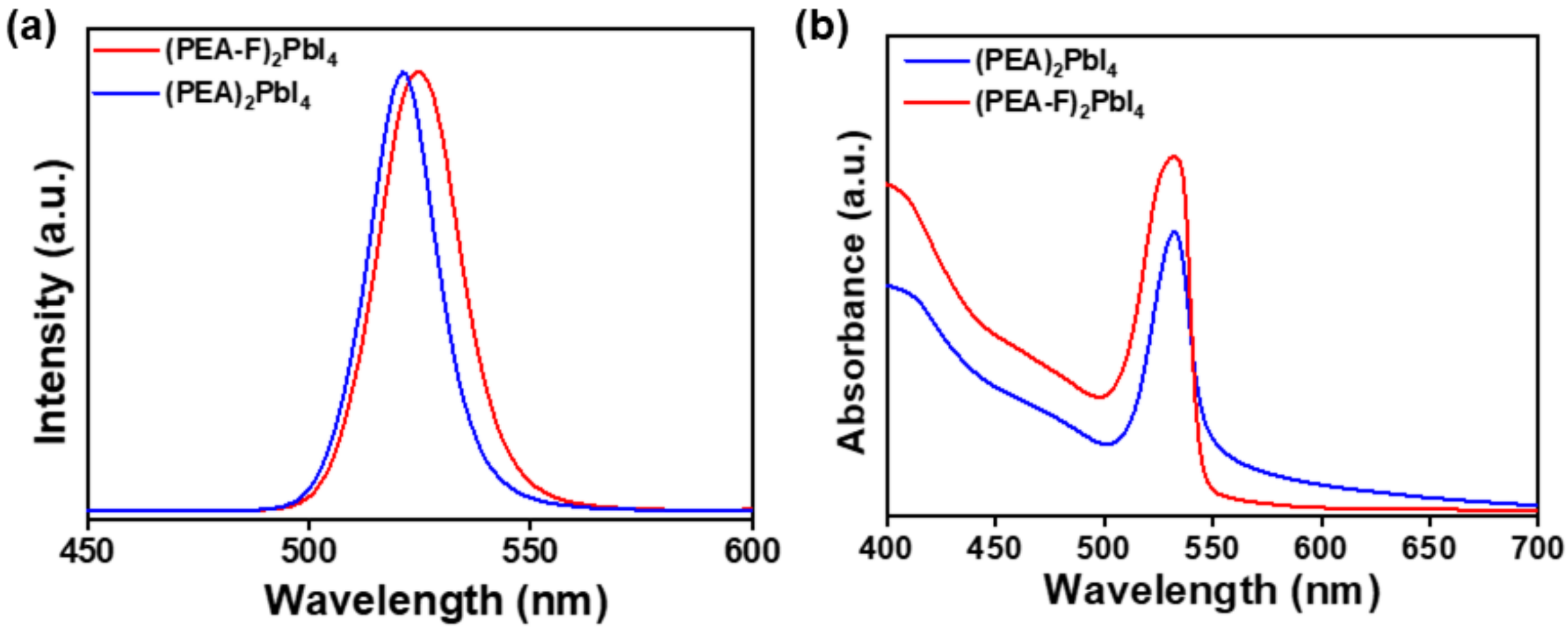

***Figure S4: Optical properties of $(PEA)_2PbI_4$ and $(PEA\text{-}F)_2PbI_4$ thin films*** *a) Normalized photoluminescence spectra of $(PEA)_2PbI_4$ and $(PEA\text{-}F)_2PbI_4$. b) Normalized absorption spectra of $(PEA)_2PbI_4$ and $(PEA\text{-}F)_2PbI_4$.*

## 2. DFT calculations:

The DFT calculations were performed using the projector-augmented wave (PAW) method, as implemented in the Vienna Ab-initio Simulation Package [1-3]. Structural optimization was performed using the Perdew, Burke, and Ernzerhof (PBE) [4] exchange-correlation functional with D3(BJ) van der Waals corrections [5]. The energy and force convergence criteria were set to $10^{-5}$ eV and 0.02 eV/Å, respectively. The plane-wave energy cutoff was set to 500 eV. Brillouin zone sampling was performed using a 2 × 2 × 1 Gamma-centered k-point grid [6]. The resulting total energies are compared in Table SI 1.

### 2.1. Phonon Calculations with Machine Learning Force Fields:

Harmonic phonon calculations were performed using the finite-displacement supercell approach described by Togo *et al.* and implemented in Phonopy [7, 8]. The phonon density of states (pDOS) was calculated using a supercell of 3×3×1 on a 21×21×7 Γ-centered q-mesh. The smearing width of $\sigma = 0.1$ was used, as implemented in the Phonopy software. Within this approach, forces on all atoms in the supercell are required to construct the force constants. For this, single-point calculations were performed. Due to the computational cost of DFT for such large systems, single-point force calculations were performed using machine-learned force fields (MLFFs). The MLFFs were trained following the protocol of Pols *et al.* [9], using the same model parameters.

The training set was generated via on-the-fly sampling [10] as implemented in VASP [1-3], based on *ab initio* molecular dynamics (MD) simulations in the NpT ensemble with a Langevin thermostat and barostat, with a friction coeffcient $\gamma = 0.5$ ps$^{-1}$. The DFT calculations employed an energy cutoff of 500 eV, an energy convergence criterion of $10^{-5}$ eV and a 2 × 2 × 1 k-point grid [6]. The SCAN [11] exchange-correlation functional was used. Details on the number of training structures and basis functions per species are provided in Table 2.

**2.2. Table S1.** Comparison of total energies (per Pb atom) for various $(PEA)_2PbI_4$ and $(F\text{-}PEA)_2PbI_4$ structures, their lattice constants and space groups after geometry optimization. The most energetically stable ones are highlighted.

| Structure | Total energy (eV) | $E_{tot}$ per Pb (eV) | a (Å) | b (Å) | c (Å) | α (°) | β (°) | γ (°) | V (Å$^3$) | Space group |
|---|---|---|---|---|---|---|---|---|---|---|
| $(PEA)_2PbI_4$ [3] | −1070.694 | −267.673 | – | – | – | – | – | – | – | – |
| **$(PEA)_2PbI_4$ [1]** | **−1070.728** | **−267.682** | **8.622** | **8.624** | **32.555** | **84.65** | **84.66** | **89.64** | **2399.655** | **P1** |
| **$(PEA\text{-}F)_2PbI_4$ [2]** | **−536.633** | **−268.317** | **8.682** | **8.518** | **16.443** | **90** | **99.53** | **90** | **1199.217** | **$P2_1/c$** |
| $(PEA)_2PbI_4$ from experimental $(F\text{-}PEA)_2PbI_4$ [2] | −534.864 | −267.432 | 8.690 | 8.525 | 16.456 | 90 | 99.53 | 90 | 1202.257 | $P2_1/c$ |
| $(PEA\text{-}F)_2PbI_4$ from experimental $(PEA)_2PbI_4$ [1] | −1072.990 | −268.248 | 8.626 | 8.627 | 32.821 | 87.42 | 87.31 | 89.38 | 2437.033 | P1 |

**2.3. Table S2.** Basis sets per atom species and number of configurations for each MLFF training set.

| System | H | C | N | F | I | Pb | Configurations |
|---|---|---|---|---|---|---|---|
| $(PEA)_2PbI_4$ | 2000 | 2000 | 507 | — | 1175 | 218 | 1030 |
| $(PEA\text{-}F)_2PbI_4$ P1 | 2000 | 2000 | 325 | 2000 | 606 | 99 | 989 |
| $(PEA\text{-}F)_2PbI_4$ $P2_1/c$ | 2000 | 2000 | 342 | 938 | 451 | 51 | 301 |

The accuracy of the MLFF-predicted forces was validated through comparison with DFT-calculated forces. The test set was constructed by randomly sampling 20 structures from a molecular dynamics (MD) trajectory, performed in the NpT ensemble at 500 K for a duration of 10 ps with a time steps of 2 fs, ensuring sufficient structural diversity. The DFT reference forces were computed using the R2SCAN exchange–correlation functional [12] with an energy cutoff of 500 eV and an energy convergence criterion of $10^{-3}$ eV, on a 2×2×1 Γ-centered k-point grid [6]. The resulting parity plots can be found in Figure 1S, showing that all three MLFFs reproduce forces close to DFT accuracy.

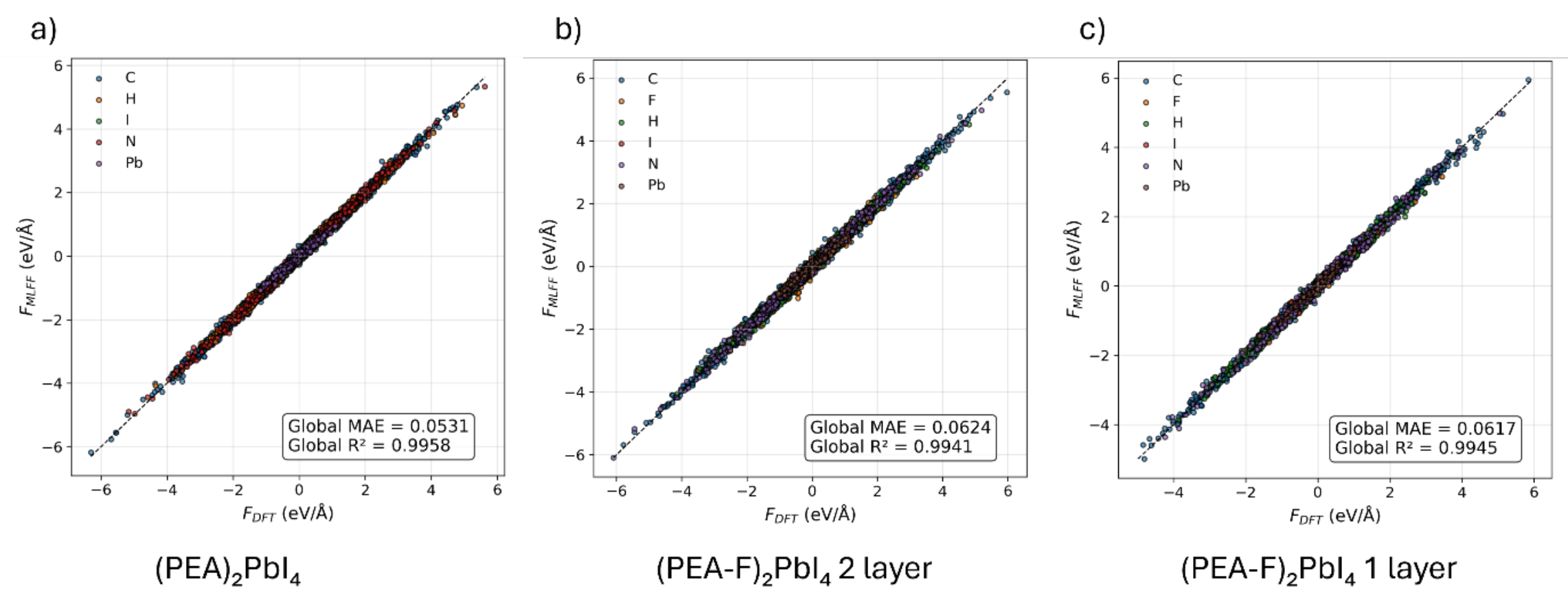

***Figure S5.*** *Parity plots comparing MLFF-predicted and DFT computed forces for all trained models. The corresponding* $R^2$ *and mean absolute error of the models are also reported.*

### 3. Comparative analysis of Raman modes

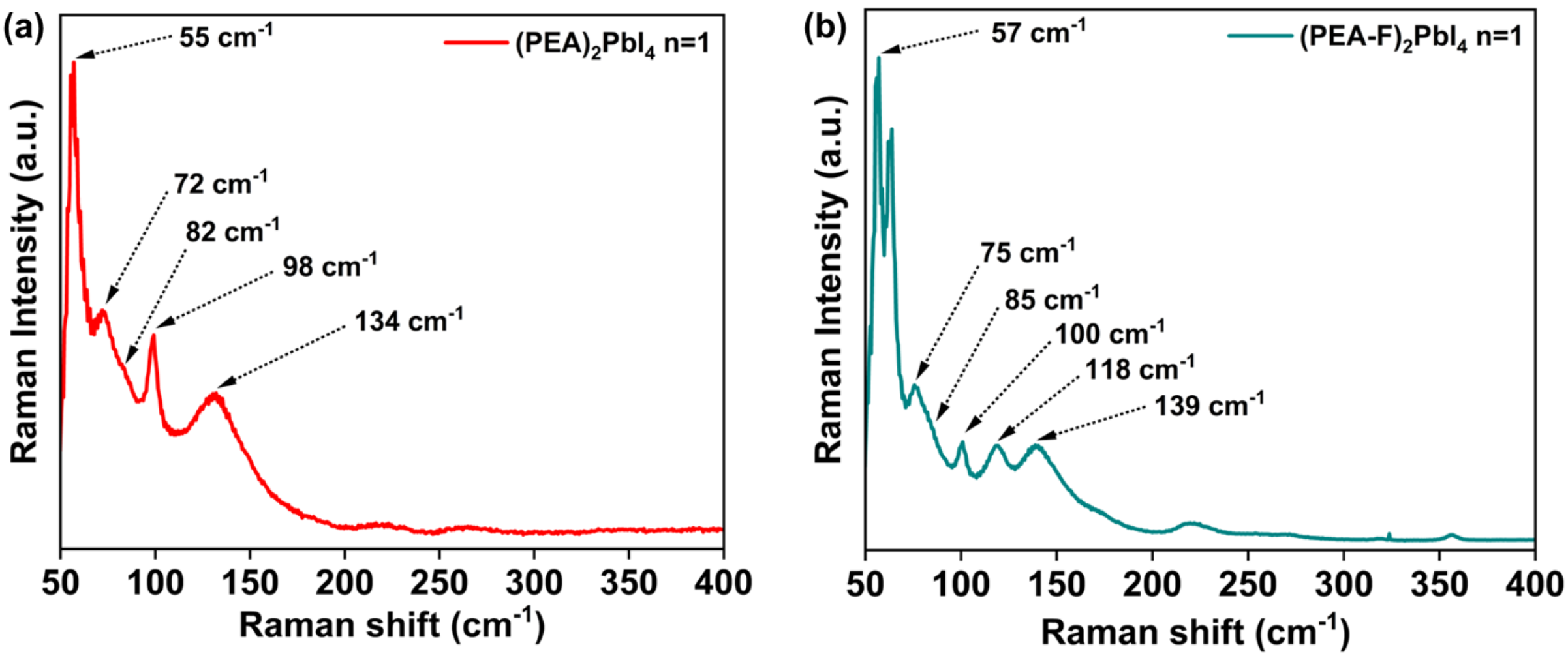

***Figure S6 presents the Raman spectra of two 2D halide perovskite materials—$(PEA)_2PbI_4$ (n=1) and $(PEA\text{-}F)_2PbI_4$ (n=1) measured at room temperature on single crystal***

*a) Raman Spectrum of $(PEA)_2PbI_4$ (n=1): This panel shows the Raman spectrum for the $(PEA)_2PbI_4$ perovskite, the spectrum displays characteristic vibrational modes associated with both the inorganic (Pb-I framework) and organic (PEA cation) components of the material. b) Raman Spectrum of $(PEA\text{-}F)_2PbI_4$ (n=1): This panel depicts the Raman spectrum for the fluorinated analogue, $(PEA\text{-}F)_2PbI_4$, where the PEA cation is modified by incorporating a fluorine atom.*

## 4. Measuring the bias-dependent Raman response of perovskite thin films:

### 4.1. $(PEA)_2PbI_4$ n=1 thin films:

To investigate the Raman response of $(PEA)_2PbI_4$ thin films, we first performed measurements at room temperature under no-bias conditions shown in Figure S7a. The device, as depicted in the Figure S7b, features a 10-micrometer channel length with pre-patterned gold electrodes on a Fraunhofer substrate. Given that our Raman spot size is approximately 1 $\mu m^2$, precise alignment was essential to ensure the laser probed the perovskite channel itself rather than the adjacent gold electrodes. For bias-dependent measurements, we applied a constant 10 V bias across the two electrodes for 120 seconds. After initiating the bias, we observed the current response of the device. Notably, as shown in Figure S7c, the current initially decreased over the first 30 seconds before stabilizing. To allow the system to reach equilibrium and eliminate transient effects, we waited for 30 seconds after applying the bias before beginning Raman data acquisition. The Raman spectra were then recorded over the subsequent 90 seconds while maintaining the applied bias. The collected Raman data from both no-bias and biased conditions were subsequently subjected to spectral fitting and analysis, enabling detailed scientific evaluation of the structural and vibrational changes induced by the applied electric field. The spectra were first baseline corrected using a fifth-order order polynomial function defined with an identical number of anchor points for all spectra. The corrected spectra were then fitted with Lorentzian functions for peak deconvolution and extraction of relevant parameters, including the full width at half maximum (FWHM). This methodological approach ensures that the observed Raman response accurately reflects the intrinsic properties of the $(PEA)_2PbI_4$ thin film under relevant electrical conditions, avoiding artifacts from electrode material and transient effects.

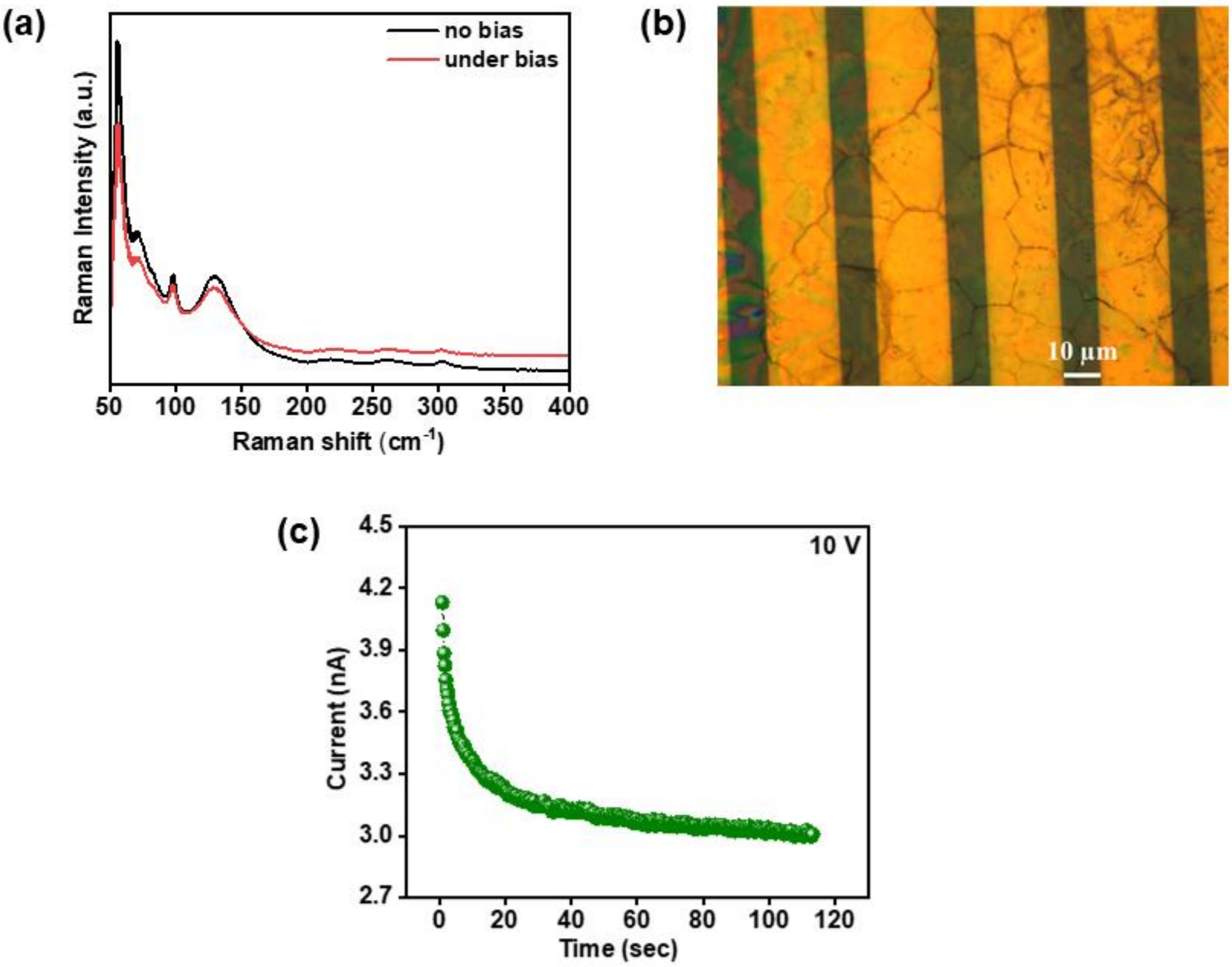

***Figure S7 presents a comprehensive analysis of the electrical and vibrational response of $(PEA)_2PbI_4$ (n=1) thin films under electrical bias, using Raman spectroscopy and current measurement:***

***(a)*** *Raman spectra of $(PEA)_2PbI_4$ (n=1) thin film under bias and no-bias conditions: displays the Raman spectra of the $(PEA)_2PbI_4$ (n=1) thin film deposited on a Si/$SiO_2$-coated substrate, recorded at room temperature (300 K) with a 633 nm excitation laser. Two sets of spectra are shown: one measured under applied electrical bias and the other without bias. Comparing these spectra allows for observation of bias-induced changes in vibrational modes.* ***(b)*** *Photograph provides a visual reference for the actual device and electrode arrangement used during the measurement.* ***(c)*** *Current response of the thin film under bias: Here, the electrical current through the $(PEA)_2PbI_4$ (n=1) thin film is plotted as a function of time while a constant 10 V bias is applied for 120 seconds. This measurement reveals the film's electrical stability,* *conduction characteristics, and possible transient phenomena such as ion migration or polarization effects during prolonged biasing.*

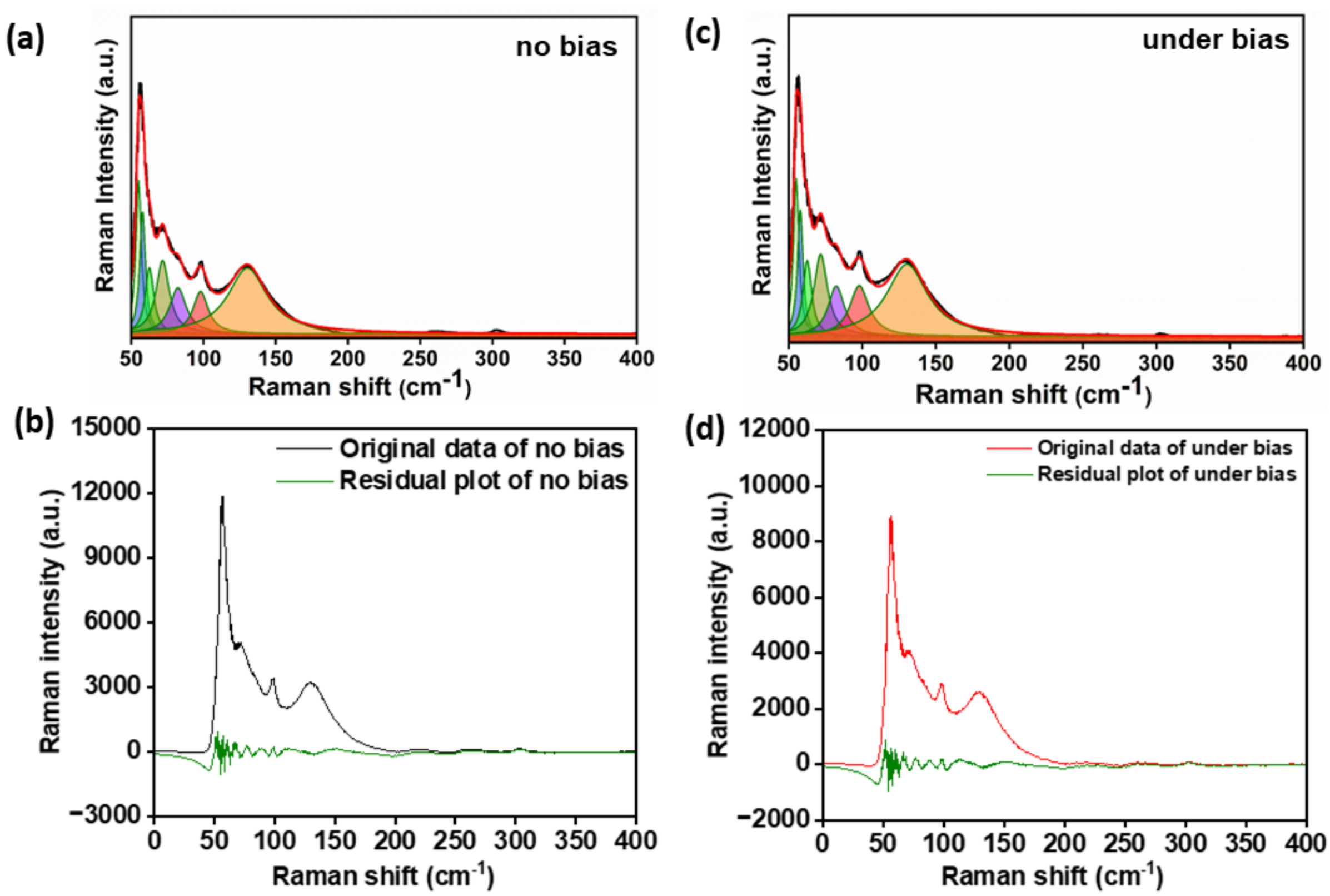

***Figure S8 provides a detailed analysis of the Raman spectral features of $(PEA)_2PbI_4$ thin films under no-bias and bias conditions by employing Lorentzian fits and evaluating the fitting accuracy through residual plots:***

***(a, c)*** *Fitting of each Raman peak under no bias and bias condition: This panel illustrates the deconvolution (fitting) of the Raman spectrum acquired from the $(PEA)_2PbI_4$ thin film under both conditions. Using a suitable Lorentzian function each observed Raman peak is individually fitted to resolve overlapping features and width/intensity of distinct vibrational modes. The peak position was fixed during the fitting procedure.* ***(b, d)*** *Residual plot after fitting under both conditions: The residual plot displays the difference between the experimental Raman data and the fitted curve for the no bias condition. Ideally, the residuals should scatter randomly around zero with minimal magnitude, indicating a good quality of fit.*

Figure S8a and Figure S8c show the fitting of the Raman spectra acquired from the $(PEA)_2PbI_4$ thin film under unbiased and biased conditions, respectively. First, the unbiased Raman spectrum was fitted using individual Lorentzian functions to resolve the spectral features and accurately determine their peak positions, full width at half maximum (FWHM), and intensities. The peak positions obtained from the unbiased spectrum were then fixed during the fitting of the biased spectrum, allowing the changes in FWHM and intensity to be evaluated independently of potential peak-position variations. This approach enables a consistent comparison of the vibrational response under applied bias, particularly for partially overlapping modes where direct spectral interpretation is challenging. The accompanying residual plot (figure S8b) provides a quantitative measure of the fitting quality. The residuals are distributed randomly about zero with low amplitude, indicating that the chosen model captures the spectral features without introducing systematic error. Any systematic variations in these parameters reflect bias-induced modifications to the lattice dynamics or local electronic environment of the material. The corresponding residuals for the biased spectrum likewise remain small and randomly distributed, confirming that the fitting approach remains robust under applied field conditions and that no additional unresolved features are introduced (figure S8d). Together, this combined fitting and residual analysis enables precise quantification of vibrational modes while providing a stringent assessment of model reliability. This approach is critical for resolving subtle, bias-induced changes in the vibrational and structural properties of $(PEA)_2PbI_4$ thin films.

**Table S3: $(PEA)_2PbI_4$ thin film Raman peak intensity comparison after Lorentzian fitting**

| **Raman shift $cm^{-1}$** | **Before Bias** | **During Bias** |
|---|---|---|
| 55 | 7249 | 5410 |
| 72 | 3472 | 2817 |
| 82 | 2178 | 1730 |
| 98 | 2004 | 1744 |
| 134 | 3112 | 2479 |

**Table S4: $(PEA)_2PbI_4$ thin film Raman peak intensity ratio comparison**

| | | |
|---|---|---|
| **Before Bias** | Stretching vs bending: ($R_1$= $I_{98}$ $cm^{-1}$/$I_{72}$ $cm^{-1}$) | 0.577 |
| | Inorganic vs organic: ($R_2$= $I_{98}$ $cm^{-1}$/$I_{134}$ $cm^{-1}$) | 0.643 |
| **During Bias** | Stretching vs bending: ($R_1$= $I_{98}$ $cm^{-1}$/$I_{72}$ $cm^{-1}$) | 0.629 |
| | Inorganic vs organic: ($R_2$= $I_{98}$ $cm^{-1}$/$I_{134}$ $cm^{-1}$) | 0.703 |

**4.2. $(PEA\text{-}F)_2PbI_4$ n=1 thin films:**

To investigate the Raman response of $(PEA\text{-}F)_2PbI_4$ thin films, measurements were first performed at room temperature under unbiased conditions (Fig. S9a). The device, shown schematically in the Fig. S9b, consists of a 20-μm channel defined by pre-patterned Au electrodes on a Fraunhofer substrate. Because the Raman laser spot is approximately 1 $\mu m^2$, precise spatial alignment was required to ensure that the laser was focused exclusively on the perovskite channel and not on the adjacent Au electrodes. For bias-dependent measurements, a constant voltage of 20 V was applied across the electrodes for 230 s. The device current was monitored upon bias application, exhibiting an initial decrease over the first ~50 s before reaching a steady state (Fig. S9c). To ensure equilibrium conditions and suppress transient effects, Raman acquisition was initiated after a 50 s delay. Spectra were then collected over the subsequent 90 s under continuous bias. The Raman spectra acquired under zero-bias and biased conditions were subsequently analyzed by spectral fitting to quantify the electric-field-induced changes in the vibrational response. All spectra were first baseline-corrected using a fifth-order polynomial with an identical number of anchor points, ensuring consistent treatment across measurements. The corrected spectra were then deconvoluted using Lorentzian functions to extract the peak positions, intensities, and full widths at half maximum (FWHM). This consistent fitting procedure enables reliable comparison of the phonon response under applied bias while minimizing contributions from spectral background and fitting-related artifacts.

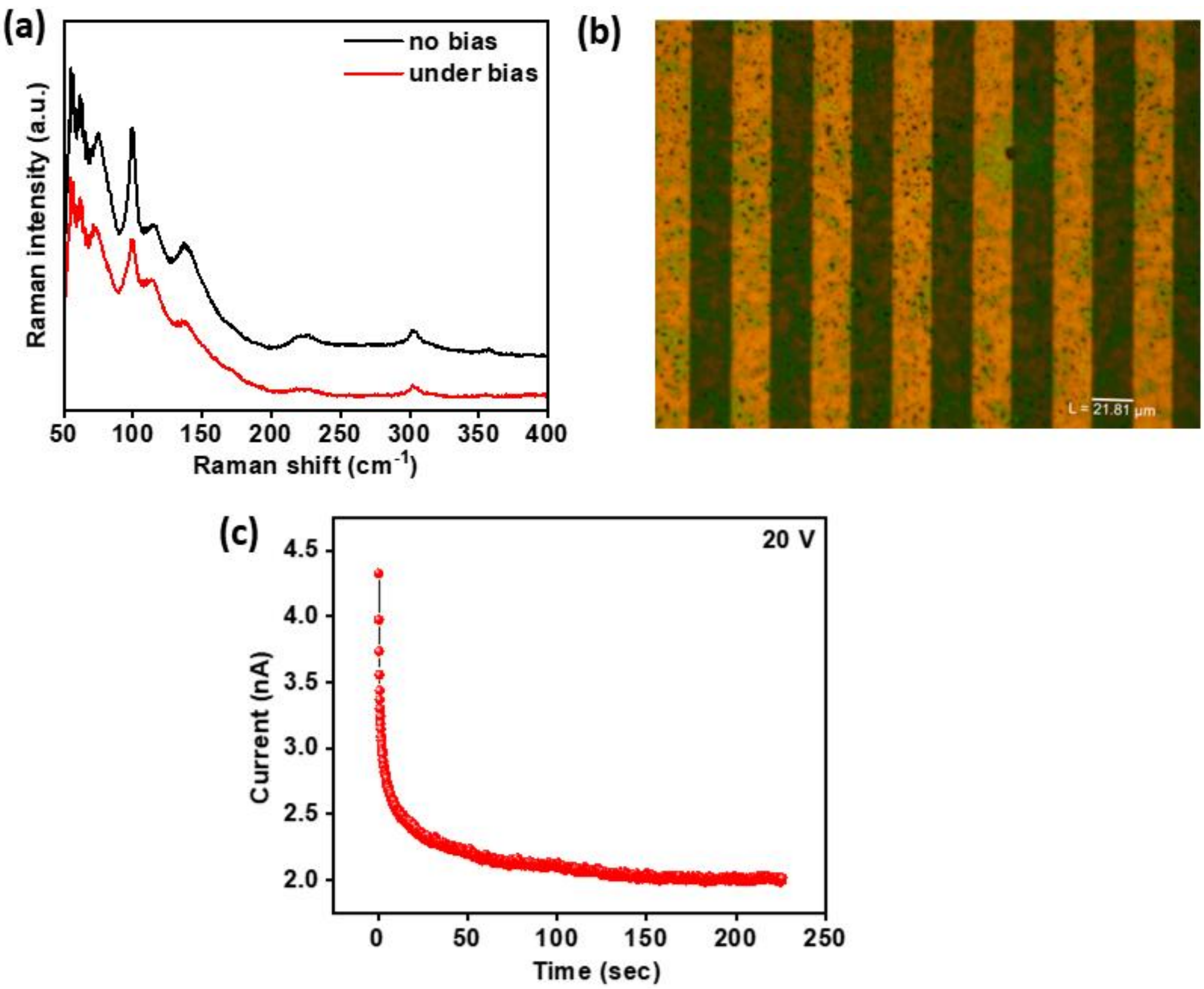

***Figure S9 presents a detailed investigation of the bias-dependent Raman response and electrical behavior of $(PEA\text{-}F)_2PbI_4$ (n=1) thin films, offering insight into the effects of an applied electric field on this fluorinated 2D halide perovskite:***

***(a)*** *Raman spectra were recorded at room temperature (300 K) using a 633 nm excitation for the thin film deposited on a $Si/SiO_2$ substrate, with measurements performed both without bias and under bias condition. Comparison of the spectra highlights the bias-induced changes in the vibrational response.* ***(b)*** *Photograph of the device, showing the thin-film channel and electrode configuration used for the Raman measurements. (c) Current response of the thin film under electrical bias. The current through the $(PEA\text{-}F)_2PbI_4$ thin film was monitored for 230 s under a constant 20 V bias, demonstrating stable electrical operation during the Raman measurement.*

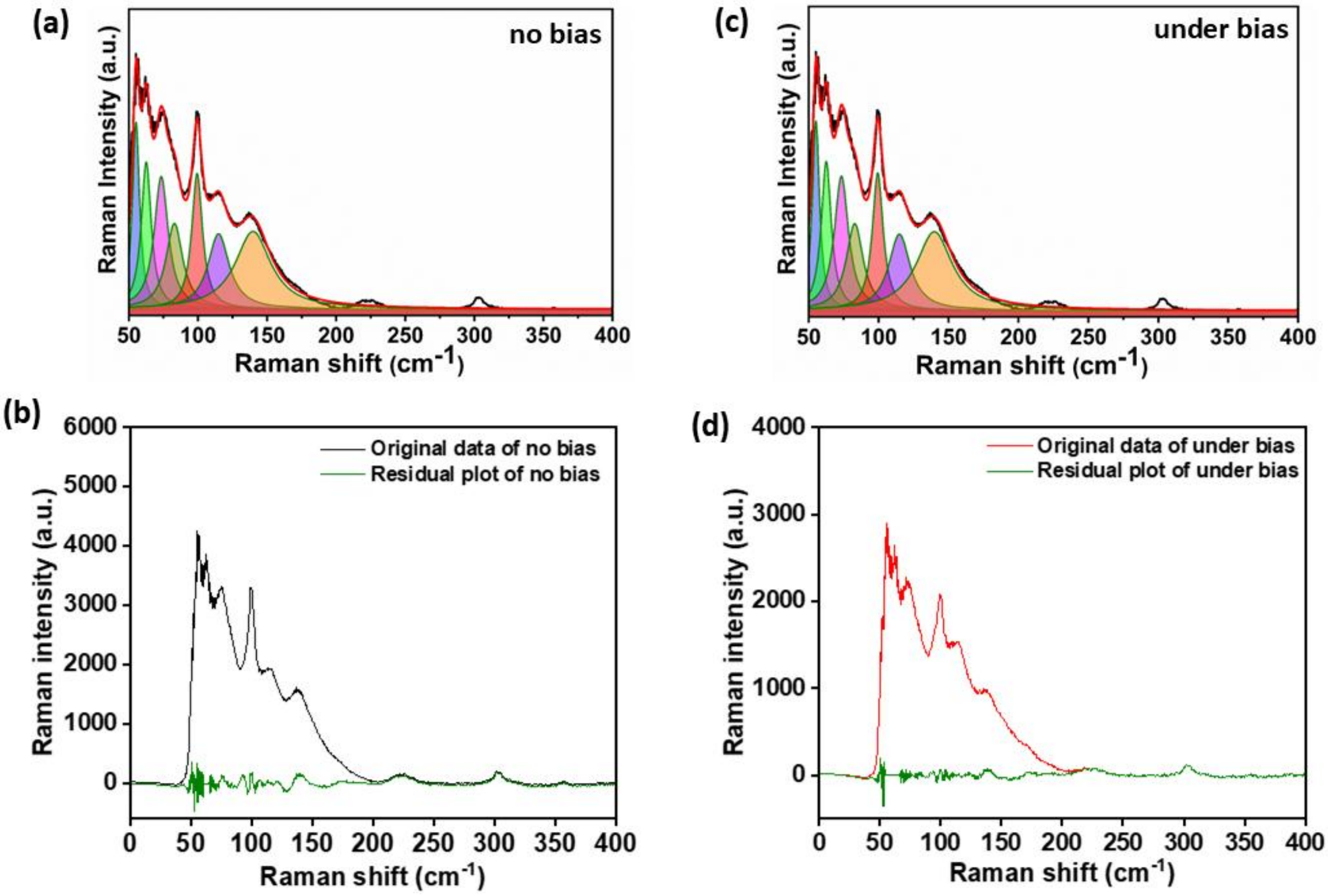

***Figure S10 provides a comprehensive analysis of the Raman spectral fitting for (PEA-F)$_2$PbI$_4$ (n=1) thin films under both no-bias and bias conditions, highlighting the accuracy of peak deconvolution and the effects of an external electric field on the vibrational features:***

***(a, c)*** *Fitting of each Raman peak under no bias and bias condition: This panel illustrates the deconvolution (fitting) of the Raman spectrum acquired from the (PEA-F)$_2$PbI$_4$ thin film under both conditions. Using a suitable Lorentzian function each observed Raman peak is individually fitted to resolve overlapping features and width/intensity of distinct vibrational modes. The peak position was fixed during the fitting procedure.* ***(b, d)*** *Residual plot after fitting under both conditions: The residual plot displays the difference between the experimental Raman data and the fitted curve for the no bias condition. Ideally, the residuals should scatter randomly around zero with minimal magnitude, indicating a good quality of fit.*

Figure S10a and Figure S10c show the Lorentzian fitting of the Raman spectra acquired from the $(PEA-F)_2PbI_4$ thin film under unbiased and biased conditions, respectively. The unbiased spectrum was first fitted using individual Lorentzian functions to resolve the vibrational features and determine their peak positions, FWHM and intensities. The peak positions obtained from the unbiased spectrum were subsequently fixed during fitting of the biased spectrum, enabling direct comparison of changes in linewidth and intensity under applied bias while minimizing uncertainties associated with peak-position variations. This approach is particularly important for resolving closely spaced and partially overlapping vibrational modes. The corresponding residuals are shown in Fig. S10b, d. In both cases, the residuals remain small and randomly distributed around zero, indicating that the fitting model adequately captures the measured spectral features without evident systematic deviations. The consistency of the residuals under both unbiased and biased conditions further supports the reliability of the fitting procedure and indicates that no additional unresolved spectral components are required. Together, the Lorentzian deconvolution and residual analysis provide a robust basis for quantifying the subtle bias-induced changes in the vibrational response of $(PEA-F)_2PbI_4$ thin films.

**Table S5: $(PEA\text{-}F)_2PbI_4$ thin film Raman peak intensity comparison**

| Raman shift $cm^{-1}$ | Before Bias | During Bias |
|---|---|---|
| 57 | 3117 | 2038 |
| 75 | 2212 | 1531 |
| 85 | 1433 | 817 |
| 100 | 2267 | 1413 |
| 115 | 1260 | 926 |
| 139 | 1300 | 763 |

**Table S6: $(PEA\text{-}F)_2PbI_4$ thin film Raman peak intensity ratio comparison**

| | | |
|---|---|---|
| **Before Bias** | Stretching vs bending: ($R_1$= $I_{100}$ $cm^{-1}$/$I_{75}$ $cm^{-1}$) | 1.02 |
| | Inorganic vs organic: ($R_2$= $I_{100}$ $cm^{-1}$/$I_{139}$ $cm^{-1}$) | 1.74 |
| **During Bias** | Stretching vs bending: ($R_1$= $I_{100}$ $cm^{-1}$/$I_{75}$ $cm^{-1}$) | 0.922 |
| | Inorganic vs organic: ($R_2$= $I_{100}$ $cm^{-1}$/$I_{139}$ $cm^{-1}$) | 1.85 |

The relative intensity ratio between the Pb–I stretching (~98 $cm^{-1}$ of $(PEA)_2PbI_4$ and ~100 $cm^{-1}$ of $(PEA\text{-}F)_2PbI_4$) and I–Pb–I bending (~72 $cm^{-1}$ of $(PEA)_2PbI_4$ and ~75 $cm^{-1}$ of $(PEA\text{-}F)_2PbI_4$)) modes provides insight into lattice dynamics. A pronounced variation in this ratio under applied bias suggests enhanced electron–phonon coupling and field-induced modulation of the Pb–I framework, likely associated with bond polarization and local strain relaxation. The variation in the relative intensity ratio between inorganic Pb–I vibrational modes and organic PEA modes provides insight into the interplay between the two sublattices. An increase in the inorganic-to-organic intensity ratio under applied bias indicates enhanced polarization and electron–phonon coupling within the Pb–I framework, suggesting that injected carriers predominantly interact with the inorganic layers. Conversely, any relative enhancement of organic modes points to increased dynamic activity or reorientation of the organic spacers, reflecting a modulation of interlayer coupling.

## 5. Measuring the bias-dependent Raman response of perovskite single crystals:

### 5.1. $(PEA)_2PbI_4$ n=1 single crystals:

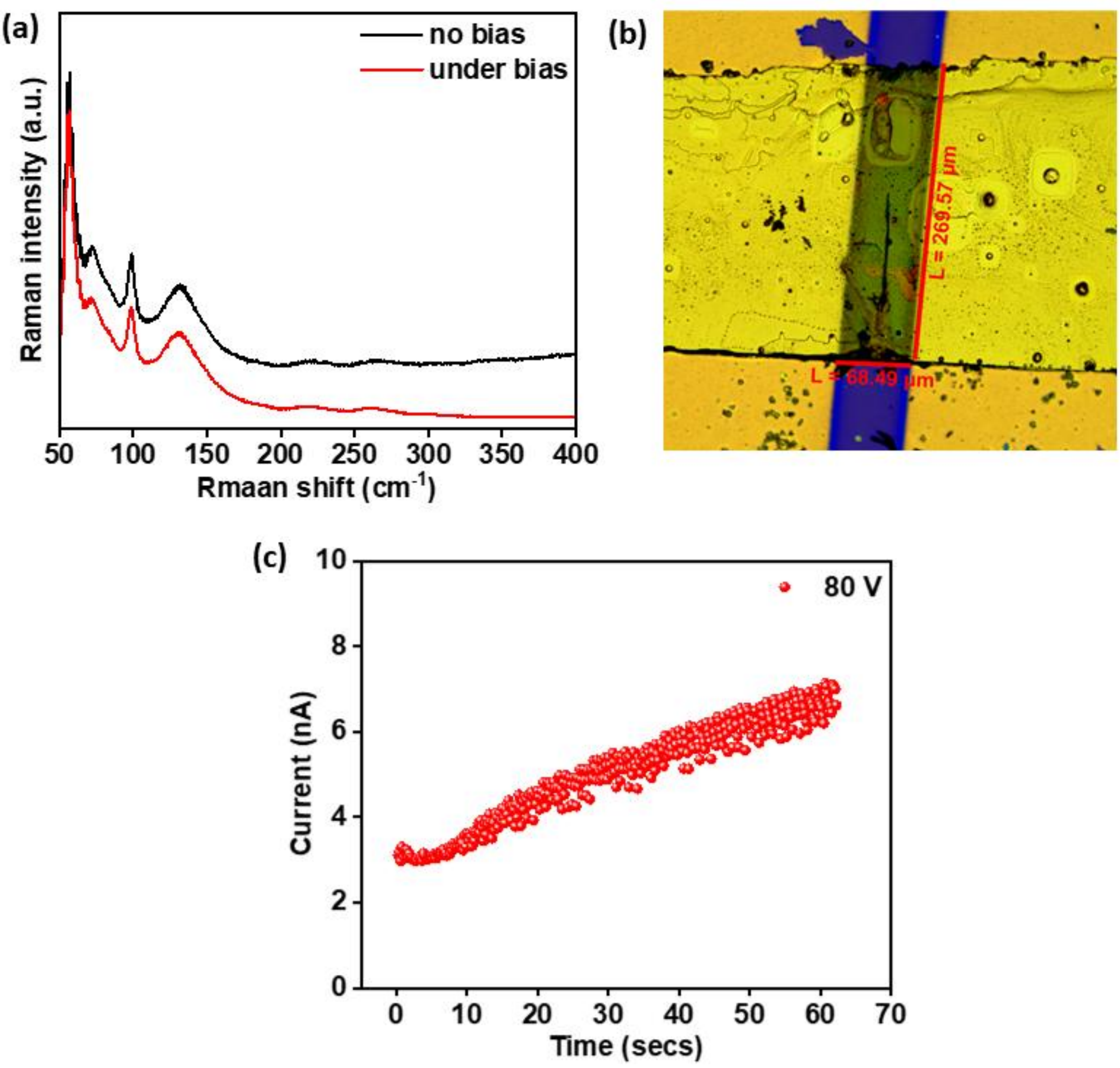

***Figure S11 Raman spectra of $(PEA)_2PbI_4$ n=1 single crystal on Si/$SiO_2$ coated substrate at 300K under bias and no bias condition excited with a laser of 633 nm:***

***a)*** *Raman spectra of a $(PEA)_2PbI_4$ (n=1) single crystal deposited on a Si/$SiO_2$ substrate, measured at 300 K using a 633 nm excitation laser. Spectra were recorded under two different electrical conditions: with an applied external bias and without bias.* ***b)*** *Optical microscopy image of the fabricated $(PEA)_2PbI_4$ single crystal device on the Si/$SiO_2$ substrate, illustrating* *the geometry and surface morphology relevant to the Raman and electrical measurements.* ***c)*** *Time-dependent current response of the* $(PEA)_2PbI_4$ *(n=1) single crystal device under a constant applied voltage of 80 V over a duration of 60 seconds. This measurement assesses the electrical stability and transport behavior of the material under prolonged bias.*

To elucidate the Raman response of $(PEA)_2PbI_4$ single crystals, we performed systematic room-temperature Raman measurements, beginning under no bias conditions (Fig. S11a). Particular care was taken to ensure precise spatial alignment of the excitation laser on the crystal surface, thereby maximizing measurement reproducibility while minimizing contributions from surface inhomogeneities or the underlying substrate. The device architecture (Fig. S11b) comprises a $(PEA)_2PbI_4$ single crystal bridging a 70 µm channel between two gold electrodes patterned on a $Si/SiO_2$ substrate. For bias-dependent measurements, a constant voltage of 80 V was applied across the electrodes for 60 s. The corresponding current–time profile (Fig. S11c) reveals a stable current throughout the biasing period. Raman spectra were subsequently collected under the applied bias over a 60 s time span, enabling direct comparison with the no bias case.

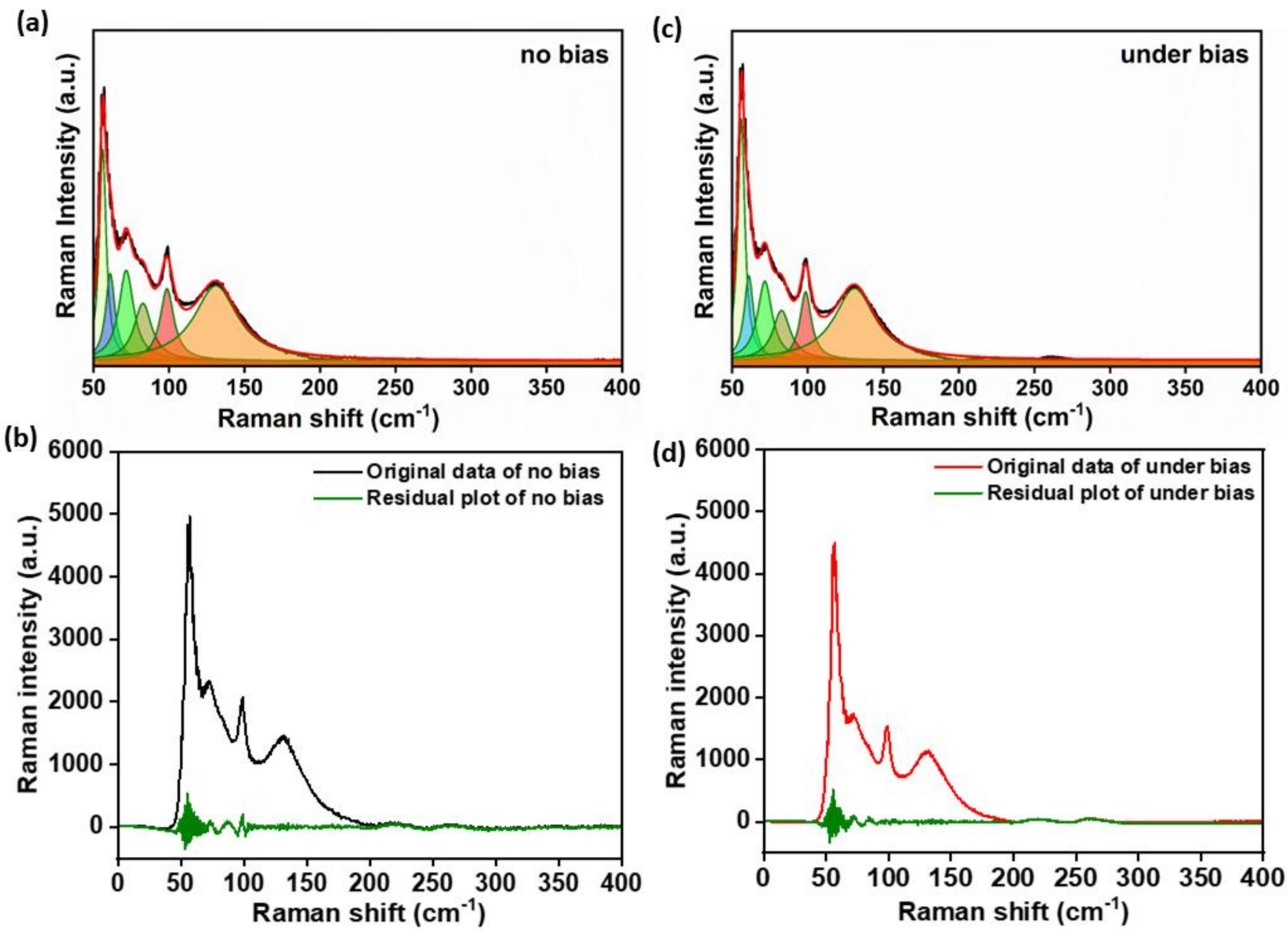

***Figure S12 provides a comprehensive analysis of the Raman spectral fitting for $(PEA)_2PbI_4$ (n=1) single crystal under both no-bias and bias conditions, highlighting the accuracy of peak deconvolution and the effects of an external electric field on the vibrational features:***

***(a, c)*** *Fitting of each Raman peak under no bias and bias condition: This panel illustrates the deconvolution (fitting) of the Raman spectrum acquired from the $(PEA)_2PbI_4$ single crystal under both conditions. Using a suitable Lorentzian function each observed Raman peak is individually fitted to resolve overlapping features and width/intensity of distinct vibrational modes. The peak position was fixed during the fitting procedure.* ***(b, d)*** *Residual plot after fitting under both conditions: The residual plot displays the difference between the experimental Raman data and the fitted curve for the no bias condition. Ideally, the residuals should scatter randomly around zero with minimal magnitude, indicating a good quality of fit.*

Raman spectra acquired under unbiased and biased conditions were analyzed by spectral fitting to quantify electric-field-induced changes in the vibrational response. All spectra were first baseline-corrected using a fifth-order polynomial with an identical set of anchor points, ensuring consistent background treatment across measurements. The corrected spectra were subsequently deconvoluted using Lorentzian functions to extract the peak positions, intensities, and full widths at half maximum (FWHM). This consistent analysis provides a reliable comparison of phonon responses under applied bias while minimizing uncertainties associated with background subtraction and spectral fitting. Figure S12a shows the fitted Raman spectrum of the $(PEA)_2PbI_4$ single crystal acquired under no bias conditions. The spectrum is fitted using Lorentzian functions, enabling the resolution of partially overlapping vibrational modes and the accurate extraction of key spectral parameters, including peak position, full width at half maximum (FWHM), and intensity. The peak positions obtained from the unbiased spectrum were subsequently fixed during fitting of the biased spectrum, enabling direct comparison of changes in linewidth and intensity under applied bias while minimizing uncertainties associated with peak-position variations. This approach is particularly important for resolving closely spaced and partially overlapping vibrational modes. The fidelity of the fit is evaluated FWHM the residuals shown in Figure S12b. These residuals, defined as the difference between the experimental spectrum and the fitted model, are randomly distributed about zero and exhibit low amplitude. An analogous fitting procedure is applied to the Raman spectrum collected under an applied electrical bias (Fig. S12c). Each Raman-active mode is parameterized using Lorentzian functions, allowing for direct and consistent comparison with the unbiased spectrum. The corresponding residuals for the biased spectrum (Fig. S12d) remain small and randomly distributed, confirming the robustness of the fitting approach under external biased.

**Table S7: $(PEA)_2PbI_4$ single crystal Raman peak intensity comparison**

| Raman shift $cm^{-1}$ | Before Bias | During Bias |
|---|---|---|
| 55 | 3839 | 3661 |
| 72 | 1641 | 1199 |
| 82 | 1041 | 755 |
| 98 | 1307 | 1038 |
| 134 | 1362 | 1091 |

**Table S8: $(PEA)_2PbI_4$ single crystal Raman peak intensity ration comparison**

| | | |
|---|---|---|
| **Before Bias** | Stretching vs bending: ($R_1$= $I_{98}$ $cm^{-1}$/$I_{72}$ $cm^{-1}$) | 0.796 |
| | Inorganic vs organic: ($R_2$= $I_{98}$ $cm^{-1}$/$I_{134}$ $cm^{-1}$) | 0.959 |
| **During Bias** | Stretching vs bending: ($R_1$= $I_{98}$ $cm^{-1}$/$I_{72}$ $cm^{-1}$) | 0.865 |
| | Inorganic vs organic: ($R_2$= $I_{98}$ $cm^{-1}$/$I_{134}$ $cm^{-1}$) | 0.951 |

**5.2. $(PEA-F)_2PbI_4$ n=1 single crystals:**

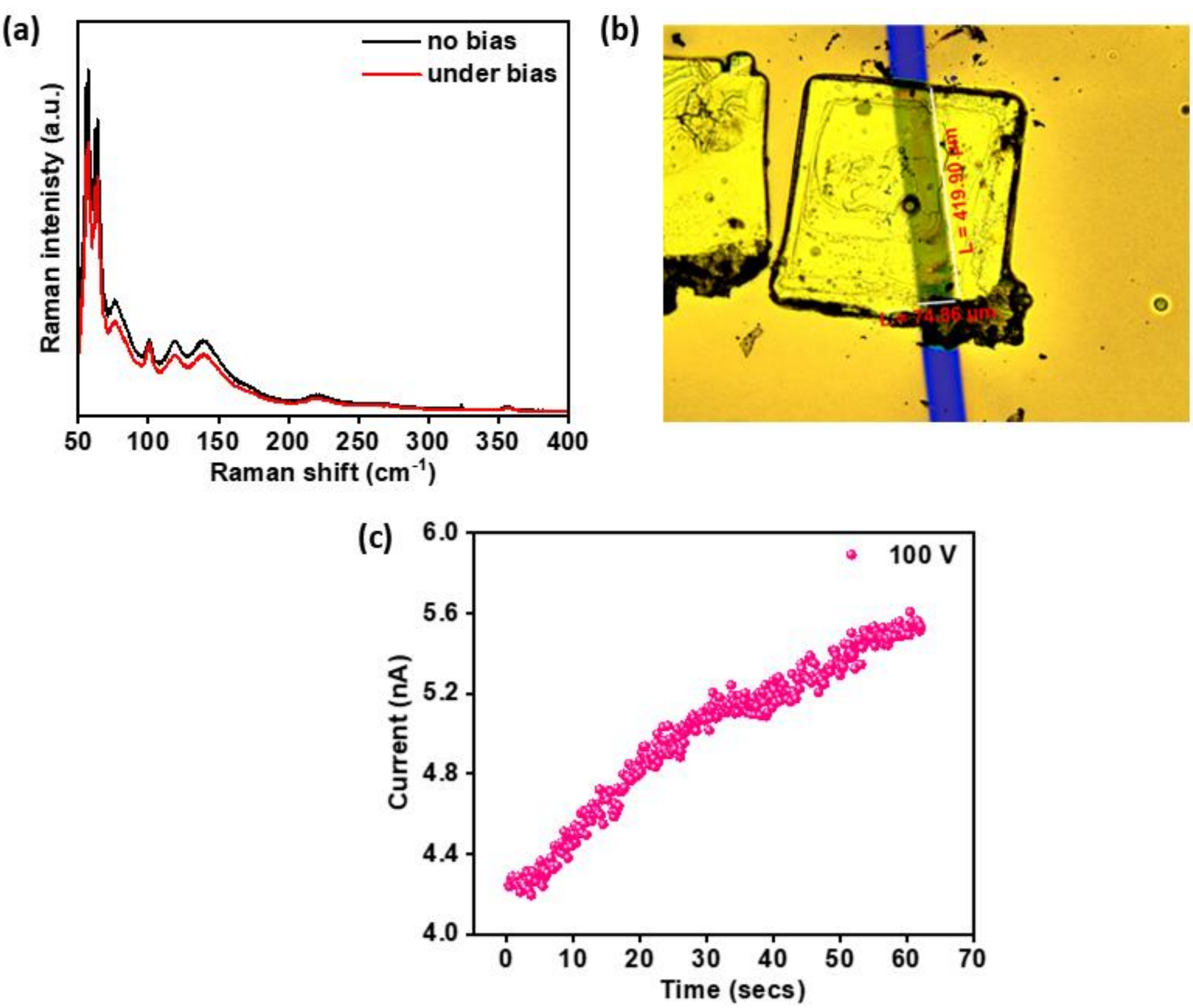

***Figure S13 presents a detailed investigation of the bias-dependent Raman response and electrical behavior of $(PEA-F)_2PbI_4$ (n=1) single crystal, offering insight into the effects of an applied electric field on this fluorinated 2D halide perovskite:***

***a)*** *Raman spectra of a $(PEA-F)_2PbI_4$ (n=1) single crystal grown on a $Si/SiO_2$ substrate, measured at 300 K using a 633 nm excitation laser. Spectra were collected under both unbiased and bias conditions.* ***b)*** *Optical microscopy image of the fabricated $(PEA-F)_2PbI_4$ single crystal device on the $Si/SiO_2$ substrate.* ***c)*** *Time-resolved current response of the $(PEA-F)_2PbI_4$ (n=1) thin film device under a constant applied voltage of 100 V for a duration of 60 seconds.*

To probe the bias-dependent vibrational response of $(PEA-F)_2PbI_4$ single crystals, room-temperature Raman measurements were first performed under unbiased conditions (Fig.

S13a). The excitation laser was carefully positioned on the crystal surface to ensure reproducible measurements while minimizing contributions from surface heterogeneity and the underlying substrate. The device consists of a $(PEA\text{-}F)_2PbI_4$ single crystal spanning a 70-µm channel, Au electrodes deposited on single crystal (Fig. S13b). For bias-dependent measurements, a constant voltage of 100 V was applied across the electrodes for 60 s. The corresponding current–time trace (Fig. S13c) remains stable throughout the biasing period, indicating stable electrical operation. Raman spectra were subsequently acquired under bias over the same 60-s interval, enabling a direct comparison of the vibrational response with the unbiased state.

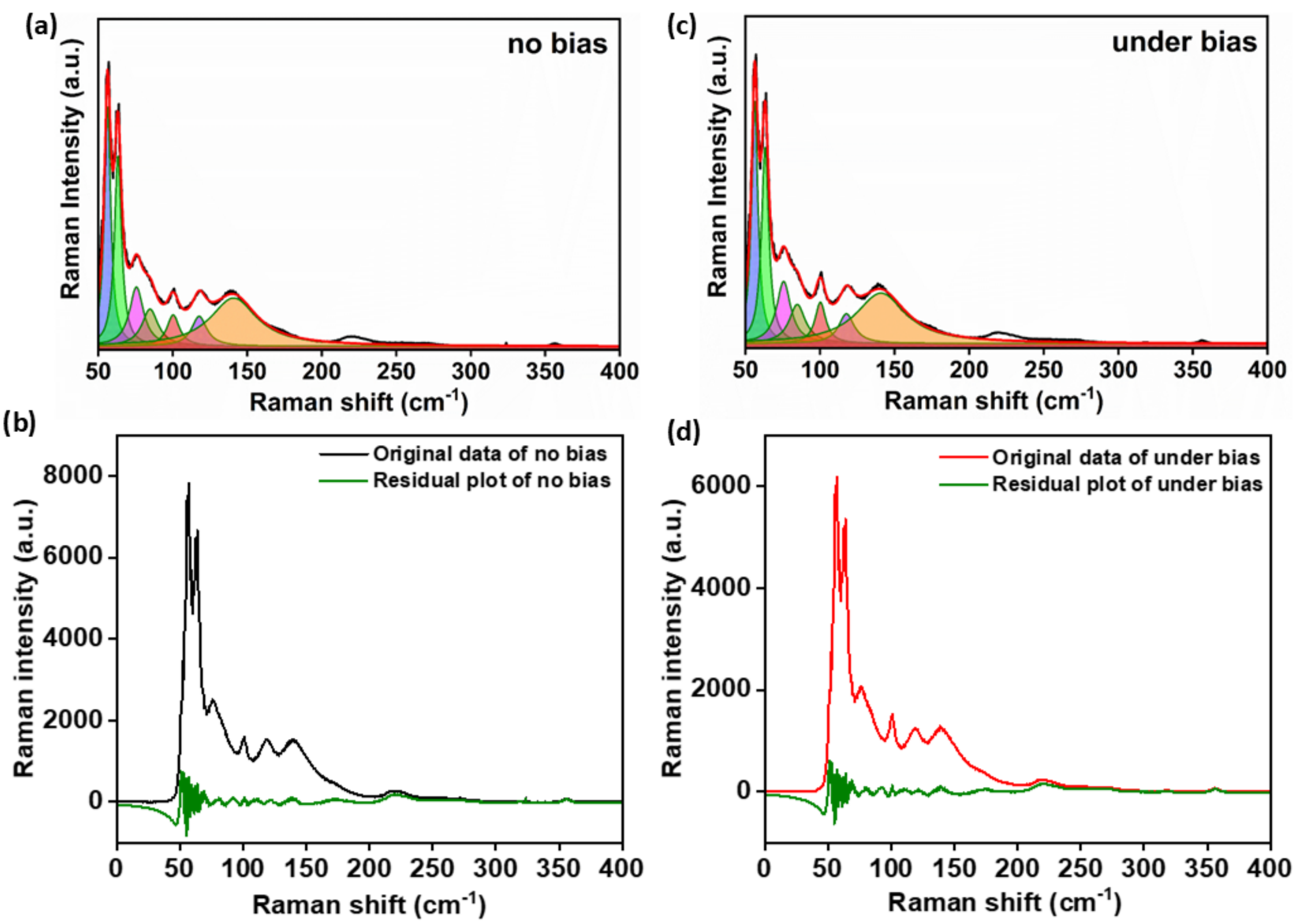

***Figure S14 provides a comprehensive analysis of the Raman spectral fitting for $(PEA\text{-}F)_2PbI_4$ (n=1) thin films under both no-bias and bias conditions, highlighting the accuracy of peak deconvolution and the effects of an external electric field on the vibrational features:***

***(a, c)*** *Fitting of each Raman peak under no bias and bias condition: This panel illustrates the deconvolution (fitting) of the Raman spectrum acquired from the (PEA-F)$_2$PbI$_4$ thin film under both conditions. Using a suitable Lorentzian function each observed Raman peak is individually fitted to resolve overlapping features and width/intensity of distinct vibrational modes. The peak position was fixed during the fitting procedure.* ***(b, d)*** *Residual plot after fitting under both conditions: The residual plot displays the difference between the experimental Raman data and the fitted curve for the no bias condition. Ideally, the residuals should scatter randomly around zero with minimal magnitude, indicating a good quality of fit.*

Figure S14a and Figure S14c show the Lorentzian fitting of the Raman spectra acquired from the (PEA-F)$_2$PbI$_4$ single crystal under unbiased and biased conditions, respectively. The unbiased spectrum was first fitted using individual Lorentzian functions to resolve the vibrational features and determine their peak positions, FWHM and intensities. The peak positions obtained from the unbiased spectrum were subsequently fixed during fitting of the biased spectrum, enabling direct comparison of changes in linewidth and intensity under applied bias while minimizing uncertainties associated with peak-position variations. This approach is particularly important for resolving closely spaced and partially overlapping vibrational modes. The corresponding residuals are shown in Fig. S14b, d. In both cases, the residuals remain small and randomly distributed around zero, indicating that the fitting model adequately captures the measured spectral features without evident systematic deviations. The consistency of the residuals under both unbiased and biased conditions further supports the reliability of the fitting procedure and indicates that no additional unresolved spectral components are required. Together, the Lorentzian deconvolution and residual analysis provide a robust basis for quantifying the subtle bias-induced changes in the vibrational response of (PEA-F)$_2$PbI$_4$ single crystal.

**Table S9: $(PEA\text{-}F)_2PbI_4$ single crystal Raman peak intensity comparison**

| Raman shift $cm^{-1}$ | Before Bias | During Bias |
|---|---|---|
| 57 | 6573 | 5178 |
| 75 | 1630 | 1332 |
| 85 | 1023 | 837 |
| 100 | 863 | 884 |
| 115 | 825 | 638 |
| 139 | 1328 | 1079 |

**Table S10: $(PEA\text{-}F)_2PbI_4$ single crystal Raman peak intensity ration comparison**

| | | |
|---|---|---|
| **Before Bias** | Stretching vs bending: ($R_1$= $I_{100}$ $cm^{-1}$/$I_{75}$ $cm^{-1}$) | 0.529 |
| | Inorganic vs organic: ($R_2$= $I_{100}$ $cm^{-1}$/$I_{139}$ $cm^{-1}$) | 0.649 |
| **During Bias** | Stretching vs bending: ($R_1$= $I_{100}$ $cm^{-1}$/$I_{75}$ $cm^{-1}$) | 0.663 |
| | Inorganic vs organic: ($R_2$= $I_{100}$ $cm^{-1}$/$I_{139}$ $cm^{-1}$) | 0.819 |

**6. Comparison of all the FWHM**

Table S11: Comparison of all the thin film and single crystals samples Raman peak FWHM at 98 cm$^{-1}$ for $(PEA)_2PbI_4$ and at 100 cm$^{-1}$ for $(PEA\text{-}F)_2PbI_4$

| Sample Name | Channel length (µm) | Voltage (V) | Electric field (V/m) | Time (secs) | FWHM | |
|---|---|---|---|---|---|---|
| | | | | | Before | During |
| **$(PEA)_2PbI_4$ thin film** | 10 | 10 | $1\times10^6$ | 90 | 12.9 ± 0.26 | 15.2 ± 0.86 |
| **$(PEA\text{-}F)_2PbI_4$ thin film** | 20 | 20 | $1\times10^6$ | 90 | 9.89 ± 0.72 | 12.75 ± 0.82 |
| **$(PEA)_2PbI_4$ single crystal** | 70 | 80 | 1.14 | 60 | 13.33 ± 0.85 | 10.67 ± 0.71 |
| **$(PEA\text{-}F)_2PbI_4$ single crystal** | 70 | 100 | $1.43\times10^6$ | 60 | 10.12 ± 0.17 | 8.40 ± 0.39 |

This table compares the full width at half maximum (FWHM) of selected Raman modes in $(PEA)_2PbI_4$ and $(PEA\text{-}F)_2PbI_4$, focusing on the characteristic peaks at 98 cm$^{-1}$ and 100 cm$^{-1}$, respectively. These low-frequency modes are particularly sensitive to changes in the structural and electronic environment, and thus serve as effective probes of lattice dynamics. For $(PEA)_2PbI_4$, the FWHM of the 98 cm$^{-1}$ mode is evaluated for both thin-film and single-crystal forms. Differences in linewidth reflect variations in crystallinity, defect density, and substrate interactions, with single crystals typically exhibiting narrower peaks consistent with reduced disorder and longer phonon lifetimes. A similar comparison is performed for the 100 cm$^{-1}$ mode in $(PEA\text{-}F)_2PbI_4$, where changes in FWHM provide insight into the influence of fluorination and morphology on lattice dynamics, including potential modifications to

electron–phonon coupling and structural coherence. More broadly, the FWHM of a Raman mode serves as a measure of phonon lifetime. Narrow linewidths indicate a well-ordered lattice with minimal scattering, whereas broader features suggest increased disorder or enhanced dynamic interactions. By comparing these parameters across compositions and sample forms, the analysis reveals how structural quality and chemical modification govern the vibrational properties and stability of layered perovskites.

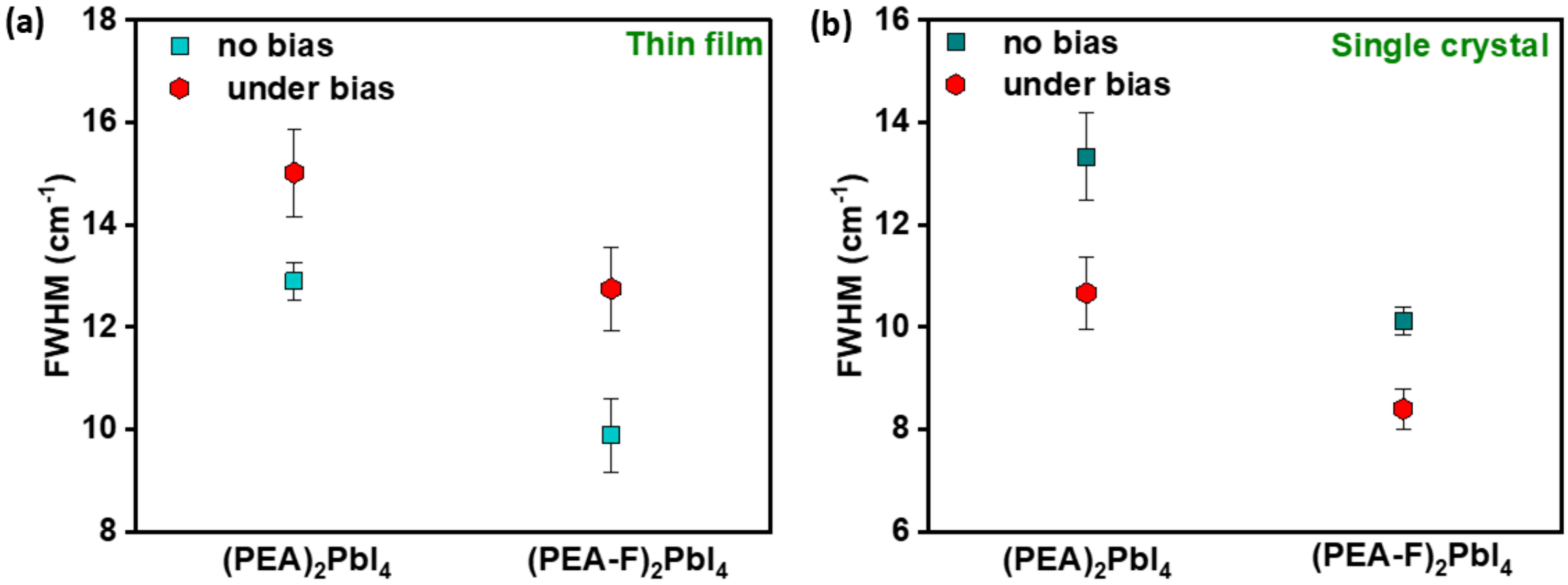

***Figure S15.*** *Calculated FWHM values for (a) the 100 cm$^{-1}$ Raman peak of $(PEA\text{-}F)_2PbI_4$ (n=1) and the 98 cm$^{-1}$ Raman peak of $(PEA)_2PbI_4$ (n=1) thin films, and (b) the 100 cm$^{-1}$ Raman peak of $(PEA\text{-}F)_2PbI_4$ (n=1) and the 98 cm$^{-1}$ Raman peak of $(PEA)_2PbI_4$ (n=1) single crystals. Error bars represent the standard error of the mean.*

Figure S15 compares the full width at half maximum (FWHM) of the low-frequency Raman modes for both thin films and single crystals of $(PEA\text{-}F)_2PbI_4$ and $(PEA)_2PbI_4$ (n = 1), providing insight into lattice dynamics and structural disorder in these materials. The FWHM values were statistically evaluated from multiple independent samples to ensure reproducibility. For $(PEA\text{-}F)_2PbI_4$, measurements were performed on five thin-film samples and four single crystals, while for $(PEA)_2PbI_4$, three thin films and three single crystals were analyzed. Error bars represent the standard error of the mean, highlighting the consistency of the measurements across multiple samples.

## 7. Comparison of phonon lifetime

**Table S12: Comparison of phonon lifetime of all the thin film and single crystals samples Raman peak at 98 cm$^{-1}$ for $(PEA)_2PbI_4$ and at 100 cm$^{-1}$ for $(PEA\text{-}F)_2PbI_4$**

| Sample Name | Channel length (μm) | Voltage (V) | Electric field (V/m) | Time (secs) | Phonon lifetime (secs) | |
|---|---|---|---|---|---|---|
| | | | | | Before | During |
| $(PEA)_2PbI_4$ thin film | 10 | 10 | $1\times10^{6}$ | 90 | 4.1 (±0.12) $\times10^{-13}$ | 3.4 (±0.19) $\times10^{-13}$ |
| $(PEA\text{-}F)_2PbI_4$ thin film | 20 | 20 | $1\times10^{6}$ | 90 | 5.4 (±0.3) $\times10^{-13}$ | 4.2 (±0.2) $\times10^{-13}$ |
| $(PEA)_2PbI_4$ single crystal | 70 | 80 | 1.14 | 60 | 3.9 (±0.49) $\times10^{-13}$ | 4.9 (±0.45) $\times10^{-13}$ |
| $(PEA\text{-}F)_2PbI_4$ single crystal | 70 | 100 | $1.43\times10^{6}$ | 60 | 5.2 (±0.15) $\times10^{-13}$ | 6.3 (±0.11) $\times10^{-13}$ |

This table presents a comparative evaluation of phonon lifetimes associated with selected Raman modes in $(PEA)_2PbI_4$ and $(PEA\text{-}F)_2PbI_4$, focusing on the characteristic peaks at 98 cm$^{-1}$ and 100 cm$^{-1}$, respectively. These low-frequency vibrations are particularly sensitive to lattice dynamics and provide insight into the structural integrity and dynamic stability of the materials. Phonon lifetimes are derived from the inverse of the full width at half maximum (FWHM) of the Raman peaks, with narrower linewidths corresponding to longer lifetimes and more coherent vibrational motion. For $(PEA)_2PbI_4$, the 98 cm$^{-1}$ mode reveals differences between thin-film and single-crystal forms, where longer lifetimes in single crystals are consistent with reduced defect densities and enhanced structural order. A similar trend is observed for $(PEA\text{-}F)_2PbI_4$ at 100 $cm^{-1}$, where variations in phonon lifetime reflect differences in crystallinity, local disorder, and electron–phonon coupling. More broadly, extended phonon lifetimes indicate reduced scattering and lower energy dissipation. By comparing phonon dynamics across material compositions and morphologies, this analysis provides valuable insight into the role of structural quality and chemical modification in governing the performance and stability of layered perovskites for optoelectronic applications.

## 8. Electrical conductivity

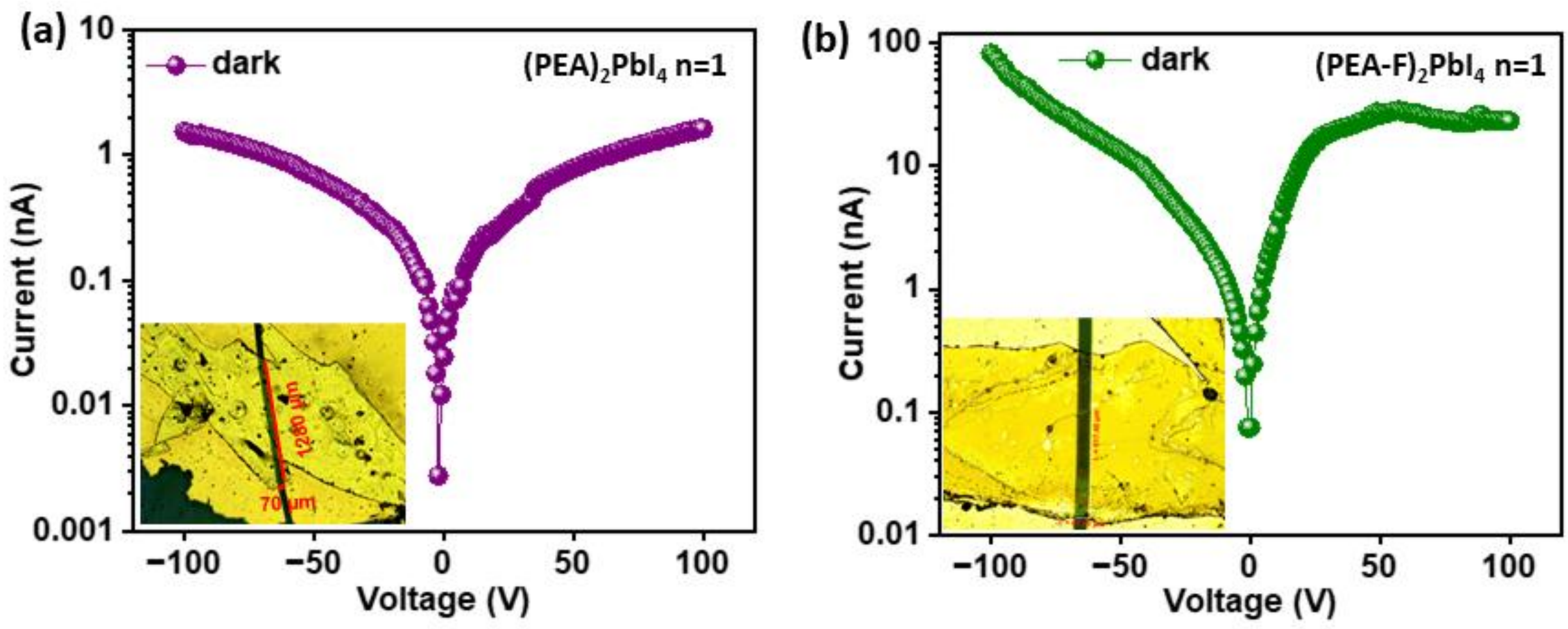

***Figure S16: Dark Current Analysis of Single Crystal Devices***

*a) Dark current of (PEA)2PbI4 based single crystal grown on Si/SiO2 substrate under 100 V, inset image shows the microscopic image of the device. b) Dark current of (PEA-F)2PbI4 based single crystal grown on Si/SiO2 substrate under 100 V, inset image shows the microscopic image of the device.*

**Table S13: Comparison of electrical conductivity of $(PEA)_2PbI_4$ and $(PEA\text{-}F)_2PbI_4$ at room temperature on single crystal**

| Sample Name | Electrical Conductivity (S/cm) |
|---|---|
| **$(PEA)_2PbI_4$** | $1.30\times10^{-9}$ |
| **$(PEA\text{-}F)_2PbI_4$** | $37.1\times10^{-9}$ |

This table compares the room-temperature electrical conductivity of $(PEA)_2PbI_4$ and $(PEA\text{-}F)_2PbI_4$ single crystals, providing insight into the impact of chemical modification on charge transport. Electrical conductivity is a key parameter governing the performance of materials in electronic and optoelectronic devices. For $(PEA)_2PbI_4$, the measured conductivity reflects the intrinsic properties of its layered perovskite structure, which are influenced by crystal quality, orientation, and defect density. In contrast, substitution with fluorinated phenylethylammonium in $(PEA\text{-}F)_2PbI_4$ alters the local electronic environment. Fluorination can modify carrier transport by influencing electronic structure, enhancing defect passivation, and potentially improving carrier mobility. The comparative conductivity values highlight how subtle compositional changes translate into measurable differences in electronic behaviour. Such variations provide insight into the interplay between structure and charge transport in two-dimensional perovskites. These findings have direct implications for device applications. Materials exhibiting higher conductivity are generally more favourable for technologies requiring efficient charge transport, including photovoltaics and light-emitting diodes. As such, this comparison offers a useful framework for assessing the suitability of chemically modified perovskites for optoelectronic applications.